%% file: ANA-PIXE-2023-02-PAPER.tex
\documentclass[PAPER, atlasdraft=false, cernpreprint, UKenglish, texmf, orcidlogo]{atlasdoc}
\usepackage{atlaspackage}
\usepackage{atlasbiblatex}

\usepackage{atlasphysics}
\graphicspath{{logos/}{figures/}}

\AtlasTitle{Sensor response and radiation damage effects for 3D~pixels in the ATLAS IBL Detector}

\AtlasAbstract{%
Pixel sensors in 3D technology equip the outer ends of the staves of the Insertable B Layer (IBL), the innermost layer of the ATLAS Pixel Detector, which was installed before the start of LHC Run 2 in 2015. 3D pixel sensors are expected to exhibit more tolerance to radiation damage and are the technology of choice for the innermost layer in the ATLAS tracker upgrade for the HL-LHC programme. While the LHC has delivered an integrated luminosity of $\simeq$\,235~fb$^{-1}$ since the start of Run~2, the 3D sensors have received a non-ionising energy deposition corresponding to a fluence of ${\simeq}\,8.5\times10^{14}$~1~MeV neutron-equivalent~cm$^{-2}$ averaged over the sensor area. This paper presents results of measurements of the 3D pixel sensors' response during Run~2 and the first two years of Run~3, with predictions of its evolution until the end of Run~3 in 2025. Data are compared with radiation damage simulations, based on detailed maps of the electric field in the Si substrate, at various fluence levels and bias voltage values. These results illustrate the potential of 3D technology for pixel applications in high-radiation environments.
}

\AtlasRefCode{PIXE-2023-02}

\PreprintIdNumber{CERN-EP-2024-156}

\AtlasJournal{Journal of Instrumentation}
\AtlasJournalRef{JINST 19 (2024) P10008}
\AtlasDOI{10.1088/1748-0221/19/10/P10008}

\hypersetup{pdftitle={ATLAS document},pdfauthor={The ATLAS Collaboration}}

\begin{document}

\maketitle

\tableofcontents

\section{Introduction}
\label{sec:intro}

%

The use of silicon pixel detectors at the LHC experiments has promoted an unprecedented R\&D effort to develop sensors and read-out electronics able to deliver the precision tracking performance required for their physics programmes while coping with the LHC beam time structure and withstanding the high radiation fluence affecting the detectors in the region close to the beam pipe.

The 3D pixel design represents one of the most innovative developments that resulted from this R\&D effort~\cite{Parker}. These novel pixels adopt a sensor geometry in which column electrodes penetrate the Si substrate instead of being implanted on the wafer surface, as in the more common planar design. The depletion region grows radially around these columns as the applied bias voltage is increased, instead of growing in depth from the collecting electrode towards the backplane. The design makes it possible to implement a sensor geometry that is intrinsically more radiation tolerant than planar sensors, by reducing the charge drift path from the point of energy deposition to the collecting electrode without reducing the thickness of the active Si material traversed by the particle.

This results in lower bias-voltage operation and higher charge collection efficiency compared to sensors of planar design, both before and after irradiation.

The Insertable B Layer (IBL)~\cite{ibl,ibltdr}, the innermost layer of the ATLAS Pixel Detector installed in the ATLAS detector~\cite{atlas-detector} before the start of LHC \RunTwo, is the first application of 3D pixels in a particle collider experiment. Now that \RunThr is well underway, it is time to assess the performance of the 3D pixel sensors, exposed to an estimated 1~\MeV neutron-equivalent (n-eq) particle fluence approaching $10^{15}$~n-eq~cm$^{-2}$. This study determines their charge collection properties and compares their performance with that of planar pixel sensors exposed to similar particle fluences. Since the 3D pixel sensors installed in the IBL detector differ in structure between a design with \enquote{fully-through} electrodes and one with \enquote{partially-through} electrodes, design-specific effects are investigated. A comparison of their responses is of interest for guiding further 3D sensor development. The time evolution of the 3D pixel response and a comparison with radiation damage simulation predictions highlight the dependence of radiation damage effects on particle fluence.

Radiation damage effects can be modelled starting from maps of the predicted electric field in the Si substrate to simulate the process of charge collection and signal generation. A radiation damage model providing these functionalities has been implemented in the official ATLAS Monte Carlo (MC) simulation software~\cite{ATL-SOFT-PUB-2021-001,ATLAS:2024vdo} and it is used to predict the detector response as a function of its operating parameters and the total particle fluence~\cite{IDET-2017-10}. Simulation results are compared with data to validate the radiation damage models and to obtain predictions for operation until the end of \RunThr.

This paper is organised as follows. The characteristics of the 3D pixels installed in the IBL are presented in Section~\ref{sec:pixels} together with a brief description of detector operations. Section~\ref{sec:raddamage} discusses the radiation exposure in \RunTwo and the first two years of \RunThr, and the modelling of the effects of radiation damage on the sensors. Section~\ref{sec:perf} presents the 3D sensor performance and its evolution with fluence in comparison with the MC predictions. Finally, Section~\ref{sec:conclusions} gives our conclusions.


\section{3D Sensors in the ATLAS Pixel Detector}
\label{sec:pixels}

%

\begin{figure}[h!]
\centering
\includegraphics[width=0.75\textwidth]{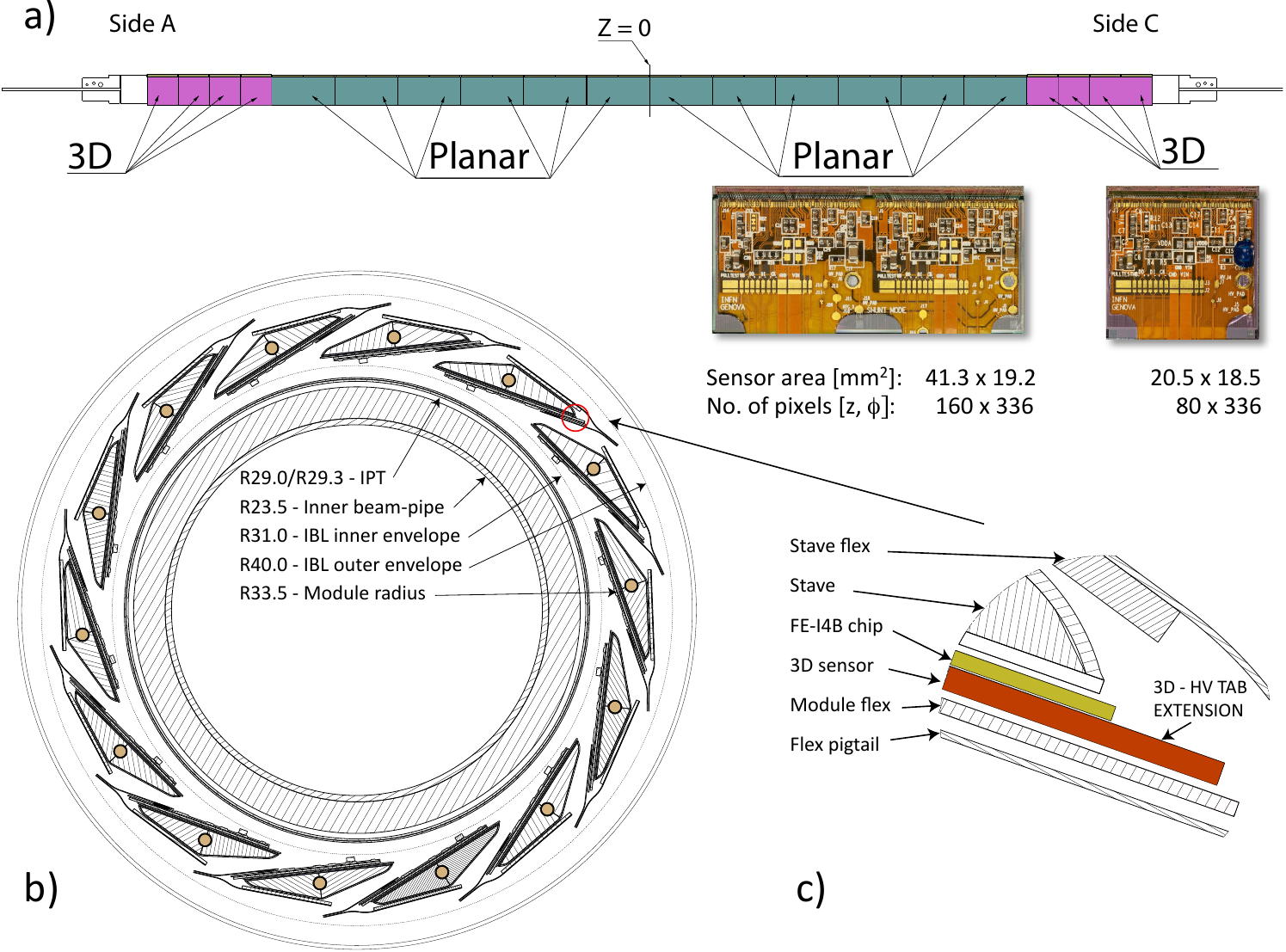}
\caption{Schematics of the IBL with the placement of the 3D modules at the ends of the staves: (a) Stave layout with the planar and 3D sensor modules. (b) Layout of the IBL with the 14 staves around the IBL positioning tube and (c) a zoomed-in view of one stave side with a 3D module visible (from Ref.~\cite{Darbo:2014kma}). \label{fig:IBL3D}}
\end{figure}

The IBL consists of 14 staves installed to form a cylinder around the LHC beam pipe at an average radius of 33.5~\si{\mm}. Each stave supports 20 pixel sensor modules together with electrical services and cooling. Two distinct sensor technologies are used: slim-edge planar sensors and 3D sensors (see Figure~\ref{fig:IBL3D})~\cite{ibl}. Planar sensors are $\sim$18~\si{\mm} wide and $\sim$41~\si{\mm} long, in the direction longitudinal to the stave. The 3D sensors have the same width and half the length. The size differences are due to manufacturing constraints related to the different processes used for sensor fabrication. All sensors have the same pixel pitch of  50~$\mu$m in the transverse coordinate and 250~$\mu$m in the longitudinal coordinate, and are read out by FE-I4 front-end (FE) chips~\cite{fei4}. The FE-I4 chip has the same overall dimensions as the 3D sensor and its channels have the same pitch as the sensors. Thus, one FE chip is bump-bonded to each 3D sensor and two FE chips are bump-bonded to each planar sensor. Sensors on each stave are organised in module groups of either four single-chip 3D sensors or two double-chip planar sensors.

Four pixel sensors in 3D technology are installed at the outer ends of the IBL staves. The IBL 3D sensors cover the longitudinal region, along the beam axis, of $245<|z|<335$~\si{mm}, in the pseudorapidity range of $1.90<|\eta|<2.55$.\footnote{ ATLAS uses a right-handed coordinate system with its origin at the nominal interaction point (IP) in the centre of the detector and the \(z\)-axis along the beam pipe. The \(x\)-axis points from the IP to the centre of the LHC ring, and the \(y\)-axis points upwards. Polar coordinates \((r,\phi)\) are used in the transverse plane, \(\phi\) being the azimuthal angle around the \(z\)-axis. The pseudorapidity, $\eta$, is defined in terms of the polar angle \(\theta\) as \(\eta = -\ln \tan(\theta/2)\).} The ATLAS Pixel Detector operates at the centre of the Inner Detector (ID), which also includes stereo pairs of silicon microstrip detector (SCT) layers and a transition radiation tracker (TRT), inside a 2~T solenoidal magnetic field~\cite{atlas-detector}.

\subsection{Design and technology}

\begin{figure}[h!]
\centering
\includegraphics[width=0.80\textwidth]{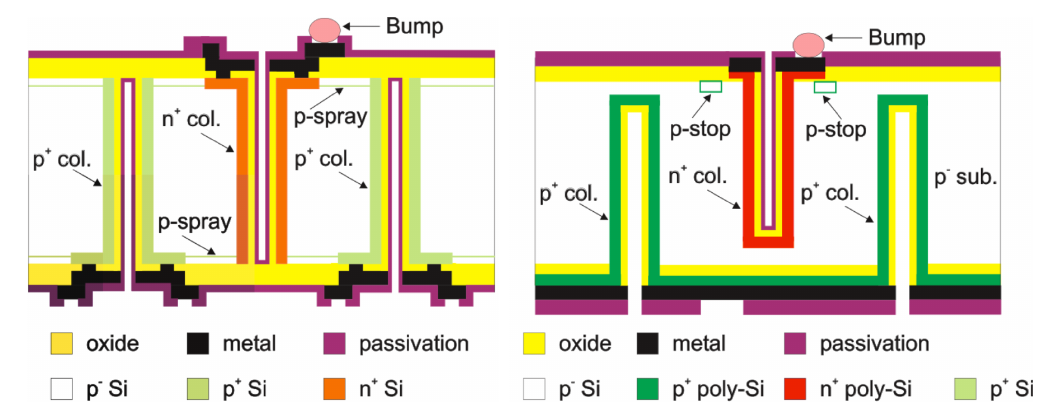}
\caption{Pixel layout and schematic cross-sections of IBL 3D sensors from FBK (left) with columns passing fully through the substrate and CNM (right), where both column types stop about 20~$\mu$m from the surface. In both cases the substrate thickness is 230~$\mu$m and the inter-electrode spacing is 67.3~$\mu$m (from Ref.~\cite{Aad:2012wf}). \label{fig:3Dschematics}}
\end{figure}

The 3D sensors installed in the IBL were manufactured to the same specifications but different designs at two production facilities: CNM, Barcelona (Spain) and FBK, Trento (Italy). Schematic cross-sections of the sensors are shown in Figure~\ref{fig:3Dschematics}.

The sensors were produced with a double-sided process~\cite{DaVia:2012ay,Aad:2012wf} on float-zone, p-type, \mbox{100-mm-diameter}, \mbox{$\langle100\rangle$-crystal-orientation}, \mbox{230-$\mu$m-thick} wafers. The bulk resistivity of the processed batch ranged between 10 and 30~k$\Omega$~cm. Columnar electrodes, which are on average 12~$\mu$m wide, were etched from opposite sides of the Si substrate using deep reactive-ion etching (DRIE) and dopant diffusion from the wafer surfaces. The n$^+$ columns and p$^+$ columns were etched from the front side and back side respectively, without using additional support wafers for mechanical stability. Thanks to this configuration, the bias voltage can be applied to all p$^+$ electrodes directly from the back side after conformal metal deposition, similarly to planar sensors.

The pixel configuration consists of two n-type readout electrodes connected at the wafer surface along the \mbox{250 $\mu$m-long} pixel direction, surrounded by six p-type electrodes, shared with the neighbouring pixels. This configuration maximises the collected charge while ensuring a minimum amount of detector capacitance and thus electronic noise. Typical values of the equivalent noise charge (ENC) measured in units of number of electrons are 130 and 140~$e^-$ for the CNM and FBK sensors, respectively, operating at a bias voltage of 20~V~\cite{Aad:2012wf}.

While the overall functionality and structure of the IBL 3D sensors processed at the two facilities are fully compatible, the designs of the FBK and CNM sensors are different. The CNM design features \enquote{partially-through}, \mbox{210 $\mu$m-long}, electrode columns, isolated on the n$^+$ side with p-stop implants~\cite{Pellegrini:2013nq}. The edge isolation is obtained with a combination of an n$^+$ grounded guard ring and fences set at the bias voltage potential from the ohmic side. The inactive edge region is about 200~$\mu$m wide. The FBK design has \enquote{fully-through} columnar electrodes, 230~$\mu$m deep, traversing the silicon bulk, isolated on both wafer sides with the p-spray technique~\cite{6522814}. A \mbox{200 $\mu$m-long} ohmic fence isolates the pixel area from the edges in the long-pitch direction. Differences between the charge collection properties of FBK and CNM sensors, due to the different column depths, were observed both before and after irradiation~\cite{DellaBetta:2013owd,Aad:2012wf}. Other differences between the two designs concern the isolation between n$^+$ electrodes and the sensor slim edge. The surface isolation between n$^+$ electrodes is obtained by implanting a p-spray layer on both wafer sides in FBK sensors, whereas p-stops are used on the front side only in CNM sensors. The slim edge is based on a multiple-ohmic-column fence able to stop the lateral spread of the depletion region in FBK sensors, while a 3D guard ring, surrounded by a double row of ohmic columns, is used to sink the edge leakage current in CNM sensors. There are 60 CNM and 52 FBK 3D sensors mounted on the IBL.

The two designs share the same top metal layout to ensure identical bump-bonding connection to the FE-I4 readout chip.  Each of the FE-I4 80$\times$336 pixels have a size of 250$\times$50~$\mu$m$^2$ and contain two readout (n$^+$) columns in the so-called 2E configuration, with an n$^+$ and p$^+$ columns' inter-electrode spacing of 67.3~$\mu$m (see Figure~\ref{fig:3DFieldPotential}). The inter-pixel cross-talk is typically below 3\%.

\subsection{Operation of IBL 3D sensors}
\label{sec:ops}

IBL 3D sensors are operated with the same power and readout systems as the IBL planar sensors. Sensors in the same module group share a common high-voltage (HV) channel and leakage current measurement. The maximum bias voltage that can be provided by the power supplies is 500~V to 3D sensors and 1000~V to planar sensors, at a maximum current of 10~mA or 8~mA respectively~\cite{ibl}.

The temperature is monitored by negative temperature coefficient (NTC) thermistors mounted on the module flex hybrid that routes signal and power lines between the stave flex hybrid board and the FE-I4 chips~\cite{ibl}. The granularity of the temperature monitoring is the same as for the HV channels. The cooling is provided by a ${\text{CO}}_2$ two-phase system, with the coolant being circulated in titanium pipes embedded in the stave structure~\cite{ibl}. The cooling was set to $-$\SI{10}{\degreeCelsius} at the start of operation in 2015, and increased to \SI{+15}{\degreeCelsius} and \SI{+5}{\degreeCelsius} in 2016 to reduce the effect of the total ionising dose (TID) on the FE-I4, which increases the leakage current of the transistors and therefore the power consumption of the chip~\cite{CYP}. After that period, cold operation was resumed with a cooling set point of $-$\SI{20}{\degreeCelsius} for most of the integrated luminosity collected during \RunTwo and \RunThr, corresponding to temperatures of $\simeq$\,${-}$\SI{13}{\degreeCelsius} measured on the modules' flex hybrids during $pp$ collision operation.  The IBL thermal history is summarised in Table~\ref{tab:IBL_Temp}.
\begin{table}
\caption{Summary of IBL temperatures (set point and module flex hybrid NTC readout) during Run~2 and Run~3. The temperature set point was changed from \SI{+15}{\degreeCelsius} to \SI{+5}{\degreeCelsius} in June 2016, after the peak of the TID effect. It was eventually lowered to $-$\SI{20}{\degreeCelsius} in 2017, when the TID effect became negligible.
\label{tab:IBL_Temp}
}
\centering
\begin{tabular}{|c|r|r|r|r|r|r|}
\hline
Temperatures & \multicolumn{4}{|c|}{Run 2} & \multicolumn{2}{|c|}{Run 3} \\
\hline
& 2015 &  2016  & 2017  & 2018  &  2022  & 2023\\
\hline
Cooling set point                        &   $-$\SI{10}{\degreeCelsius} &  \SI{+15}{\degreeCelsius}; ~~\SI{+5}{\degreeCelsius} & $-$\SI{20}{\degreeCelsius} &  $-$\SI{20}{\degreeCelsius} & $-$\SI{20}{\degreeCelsius} & $-$\SI{20}{\degreeCelsius} \\
\hline
Average module NTC readout   &    $-$\SI{3}{\degreeCelsius} &   \SI{+19}{\degreeCelsius}; \SI{+10}{\degreeCelsius}  &  $-$\SI{13}{\degreeCelsius}&   $-$\SI{13}{\degreeCelsius} & $-$\SI{13}{\degreeCelsius} & $-$\SI{13}{\degreeCelsius}\\
\hline
\end{tabular}
\end{table}
The bias voltage applied to the 3D sensors was increased progressively with time: from 20~V in 2015 to 40~V in 2017 in \RunTwo, and to 60~V in 2022 at the start of \RunThr, to ensure that the sensors were operated above the depletion voltage. Regular I--V scans were performed during periods without beam to monitor the evolution of breakdown voltages with accumulated radiation. At the end of 2023, most of the sensors show a breakdown voltage higher than 80~V. Only four HV channels show an early breakdown, while three exhibit an anomalous drifting current during $pp$ collisions. As a result, 7 channels out of 28 had to be operated at bias voltages lower than the nominal operating value.

\begin{figure}[h!]
\centering
\subfloat[]{
\includegraphics[width=0.495\textwidth]{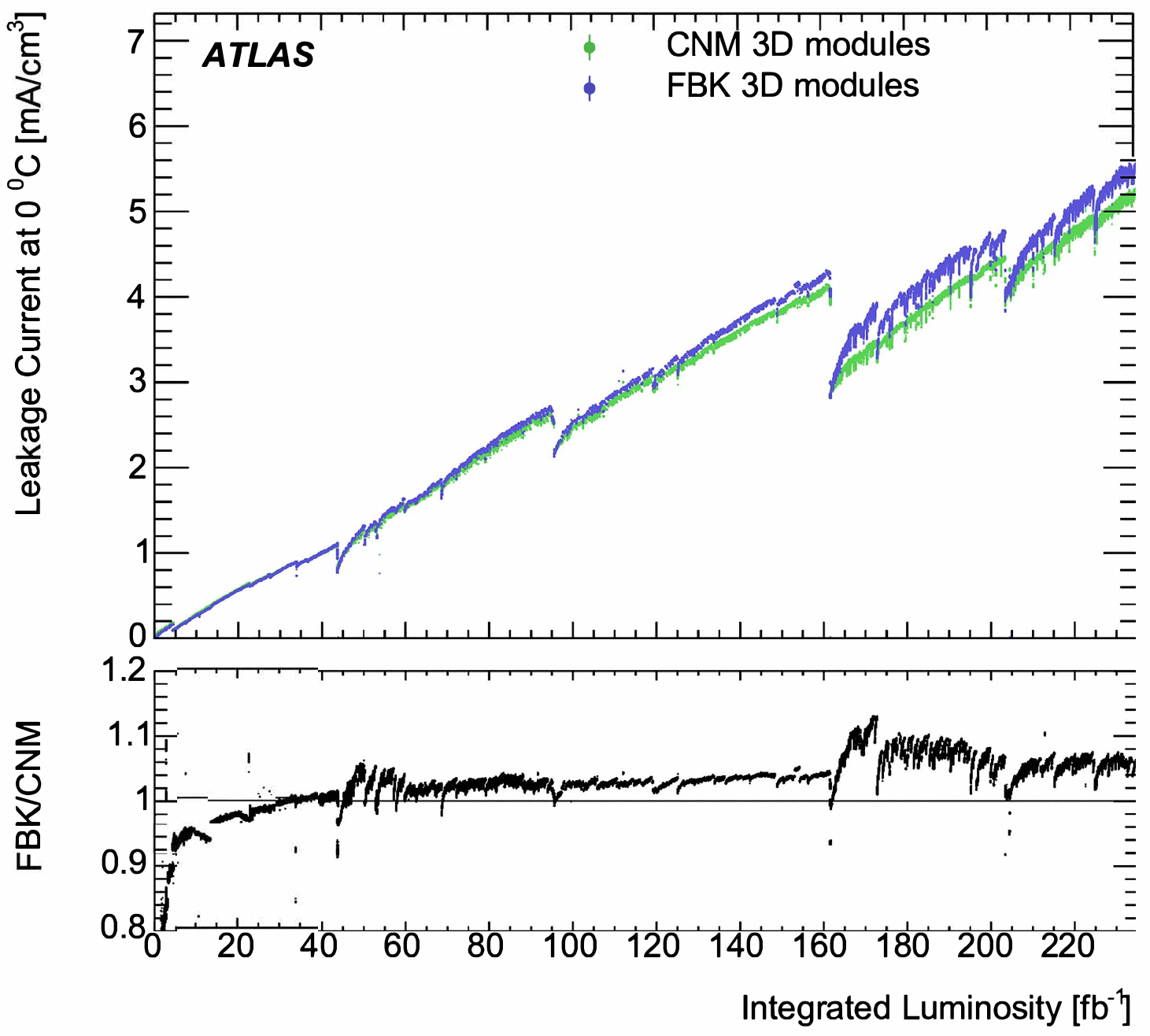}
\label{fig:IBL_FBKvsCNM}
}
\subfloat[]{
\includegraphics[width=0.495\textwidth]{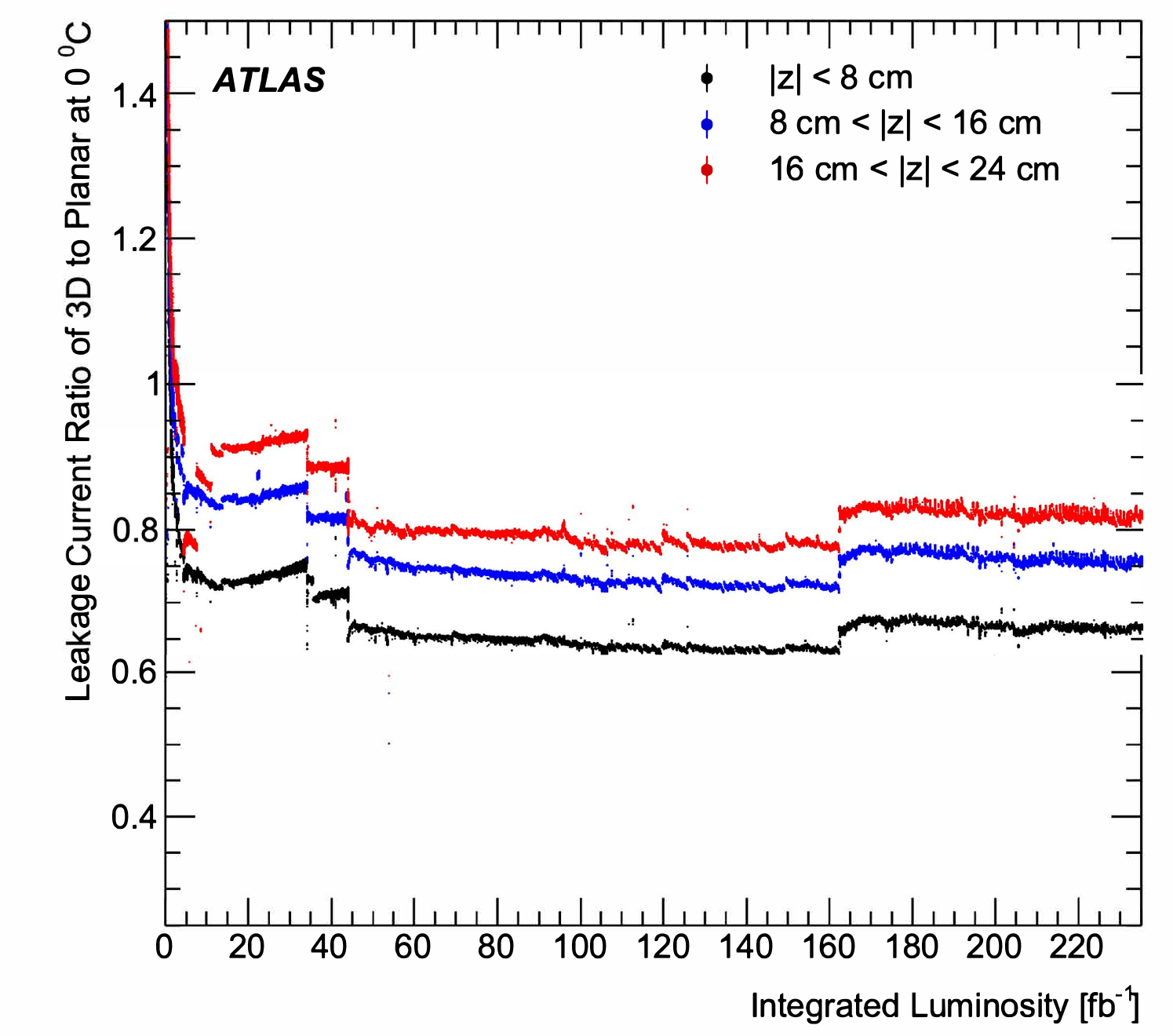}
\label{fig:IBL_Ratio}
}
\caption{ \protect\subref{fig:IBL_FBKvsCNM} Measured leakage currents (normalised to \SI{0}{\degreeCelsius}) for IBL sensors in module groups populated with 3D sensors of CNM and FBK design during \RunTwo and \RunThr and their ratio. \protect\subref{fig:IBL_Ratio} Ratios of measured leakage currents (normalised to \SI{0}{\degreeCelsius}) for IBL 3D sensors ($24<|z|<32$~cm) to those for planar sensors in three regions along $z$ during \RunTwo and \RunThr. Sudden variations observed in the ratio correspond to changes in the operating bias voltage. HV channels with sensors at bias voltages lower than the nominal value were excluded from the measurements. The different levels of leakage current observed for the planar sensors reflect the different levels of irradiation fluence along the IBL staves.}
\end{figure}

Leakage current measurements for 3D sensors show the expected linear increase with accumulated radiation damage. This increase is proportional to the fluence received and the depleted volume. Figure~\ref{fig:IBL_FBKvsCNM} shows the leakage current measured separately for the 3D detectors of FBK and CNM design, over the \RunTwo and \RunThr periods, with the expected drops during the non-irradiation annealing periods (technical stops, winter shutdowns, and Long Shutdown~2, LS2, which is visible at ${\sim}160$ fb$^{-1}$ of integrated luminosity).

The differently designed 3D sensors exhibit different responses to high irradiation and instantaneous luminosity. FBK sensors have higher rate of leakage current increase with fluence than CNM sensors, but they also show a stronger annealing effect, making the ratio of their leakage current to that of CNM sensors decrease significantly during the non-irradiation annealing periods and raise at the restart of the LHC collision periods.

Changes in bias voltage and temperature strongly affect the evolution of the detector leakage currents. This can be observed in the trend of the ratio of leakage currents for 3D ($24<|z|<32$~cm) and planar ($|z|<8$~cm, $8<|z|<16$~cm, and $16<|z|<24$~cm) sensor module groups shown in Figure~\ref{fig:IBL_Ratio}. These measurements span the whole of \RunTwo and the first two years of \RunThr. Data from all HV channels were normalised to \SI{0}{\degreeCelsius}, following the procedure discussed in Ref.~\cite{IDET-2020-01}, before being averaged over the module groups with the same $|z|$ coordinate. Steps in the ratio are visible every time the bias voltages were adjusted in the 3D or planar sensors (typically during the winter shutdowns). However, a linear increase in the ratio of 3D sensor leakage current to planar sensor leakage current was observed during 2016 for fixed values of the bias voltages, at integrated luminosity values in the range from 10 to 35~fb$^{-1}$. This can be explained by planar sensors not being fully depleted when operating at 80~V in 2016. The bias voltage applied to the IBL planar sensors was therefore increased to 150~V in September 2016 (after 35~fb$^{-1}$), for the second part of the 2016 data-taking period, and to 350~V at the start of the 2017 data taking (after 45~fb$^{-1}$).
During the long irradiation periods, when the voltages are kept constant, the ratio of the 3D to planar sensor leakage currents shows smooth behaviour.


\section{3D Sensors, Fluence and Radiation Damage Simulation}
\label{sec:raddamage}


%

\subsection{3D pixel radiation damage}

The signal amplitude produced by a minimum-ionising particle in a silicon sensor is due to the currents induced in the electrodes by electrons drifting towards them and holes moving away, following the theorem proposed by Shockley~\cite{doi:10.1063/1.1710367} and Ramo~\cite{Ramo}. After irradiation, the concentration of stable defects in the Si bulk contributes to a linear increase of leakage current, an increase of space charge, and an increase of the trapping probability for both electrons and holes. While the leakage current and the incremented space charge can be controlled by cooling or by increasing the bias voltage respectively, charge trapping, which is usually predominant after heavy irradiation, is more difficult to control.
The interplay between the signal generated by the drifting carriers and its reduction after irradiation, due to the presence of trapping sites, is modelled by applying the Shockley--Ramo theorem. A strong electric field reduces the time taken by the charge carriers to reach the collecting electrode, making it shorter than the average trapping time.

Pixels in 3D technology are known to be more tolerant to bulk radiation damage because the drift path and collection time for charge carriers is shorter. Simulation indicated that the average charge-collection time for electrons in sensors exposed to $10^{15}$~n-eq~cm$^{-2}$ is 0.56~ns for the IBL 3D sensors and 0.90~ns for IBL planar sensors, at typical operating bias voltages. The radiation tolerance was shown to depend on the column inter-electrode spacing (IES), with smaller IES values corresponding to a smaller loss of charge collection efficiency after irradiation~\cite{DAVIA2009319}. This is particularly true for sensor performance after high radiation exposure. A review of radiation tolerance of 3D pixel sensors and its dependence on their exact geometry can be found in Ref.~\cite{davia2019}. However, most of the studies conducted so far on these sensors have focused on relatively large fluences, those expected at the IBL detector's end-of-life with the completion of \RunThr and above. This study analyses the evolution of the sensor response, measured in collision data and predicted by simulation, since the start of operation in 2015.

\subsection{Radiation levels for the IBL 3D sensors}

An estimate of the particle fluence affecting the IBL detectors is an essential input for simulating the radiation damage conditions and the sensor response corresponding to a given period of LHC data taking. In the present analysis, this is achieved by studying the sensor leakage current during \RunTwo to derive the rates of particle fluence per unit of delivered integrated luminosity as a function of the radial position of the sensor relative to the beam axis and the longitudinal ($z$) position of energy deposition on its surface~\cite{IDET-2020-01}. Given the position of the IBL 3D sensors at the ends of the stave, the fluence profile along the longitudinal coordinate is particularly important. The leakage current analysis has shown the fluence's longitudinal profile to be less uniform than the FLUKA~\cite{BATTISTONI201510} and \GEANT~\cite{Agostinelli:2002hh} simulation predictions~\cite{IDET-2020-01}. In fact, the fluence delivered to the IBL 3D sensors is measured to be $\sim$\,30\% lower than that registered on the central modules equipped with planar sensors. The conversion factor between delivered integrated luminosity and the fluence on 3D sensors is obtained by averaging values obtained as a function of the $z$ position weighted by the $z$ distribution of measured particle-track impacts on the IBL surface in the data samples used for the analysis. This conversion factor is $(3.6 \pm 0.4) \times 10^{12}$~n-eq~cm$^{-2}$/fb$^{-1}$, where the quoted uncertainty accounts for systematic uncertainties from the leakage current analysis discussed in Ref.~\cite{IDET-2020-01}.

During \RunTwo, the 3D sensors received an estimated average fluence of $5.8 \times 10^{14}$~n-eq~cm$^{-2}$, with an additional contribution of $2.6 \times 10^{14}$~n-eq~cm$^{-2}$ in the first two years of \RunThr LHC operations. The total fluence is expected to approach $1.5 \times 10^{15}$~n-eq~cm$^{-2}$ by the end of \RunThr in 2025.

\subsection{Radiation damage simulation}

The ATLAS Collaboration has developed a detailed simulation of radiation damage effects in the pixel sensors~\cite{IDET-2017-10}. Signals induced on the read-out electrodes are obtained from a detailed simulation of the drift of charge carriers produced by ionising particles, taking into account radiation damage effects. The carriers' speed is calculated using the product of their mobility~\cite{JACOBONI197777} and the estimated electric field in the Si bulk. This is simulated using TCAD\footnote[3]{Technology Computer Aided Design} tools. Effective defects are used in the simulation to reproduce the effects of radiation damage.

The Chiochia radiation damage model~\cite{CHIOCHIA2006} is used to simulate the electric field profile in planar sensors after irradiation. This profile is used to calculate the average charge deflection due to the Lorentz force in the ATLAS solenoidal magnetic field. The combined effect of the electric and magnetic fields is used to estimate the time for a carrier to reach the respective collecting electrode. If the expected collection time is larger than an exponentially distributed trapping time with average value equal to $\tau = 1/(\beta \Phi)$, where $\beta$ is the trapping constant and $\Phi$ the fluence, then the carrier is considered trapped and its induced signal is calculated using a precomputed weighting potential~\cite{Ramo} map of the sensor.

\begin{figure}[ht]
\centering
\subfloat[]{
\includegraphics[width=0.425\textwidth]{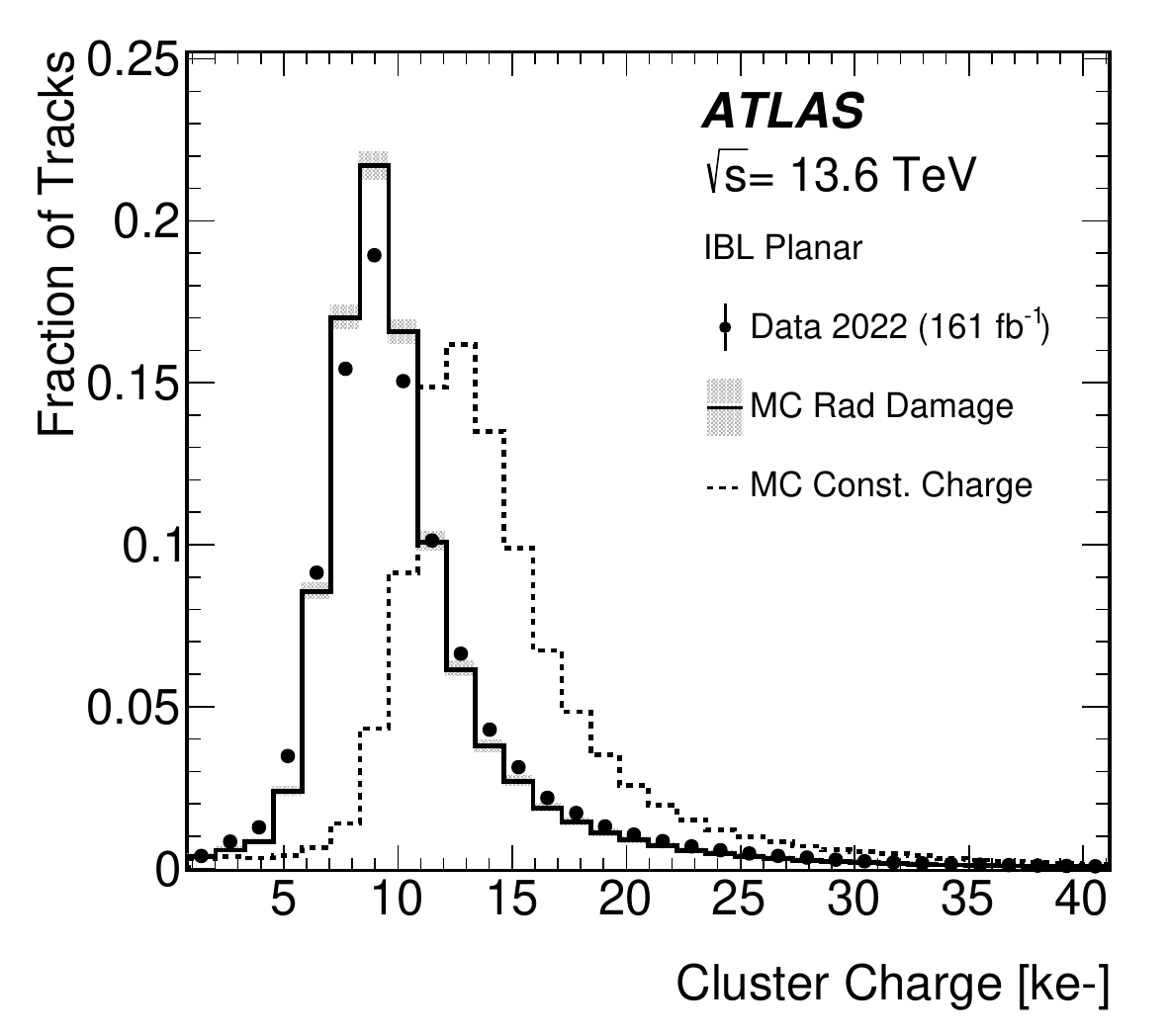}
\label{fig:Planar_Cluster_Charge}
}
\subfloat[]{
\includegraphics[width=0.475\textwidth]{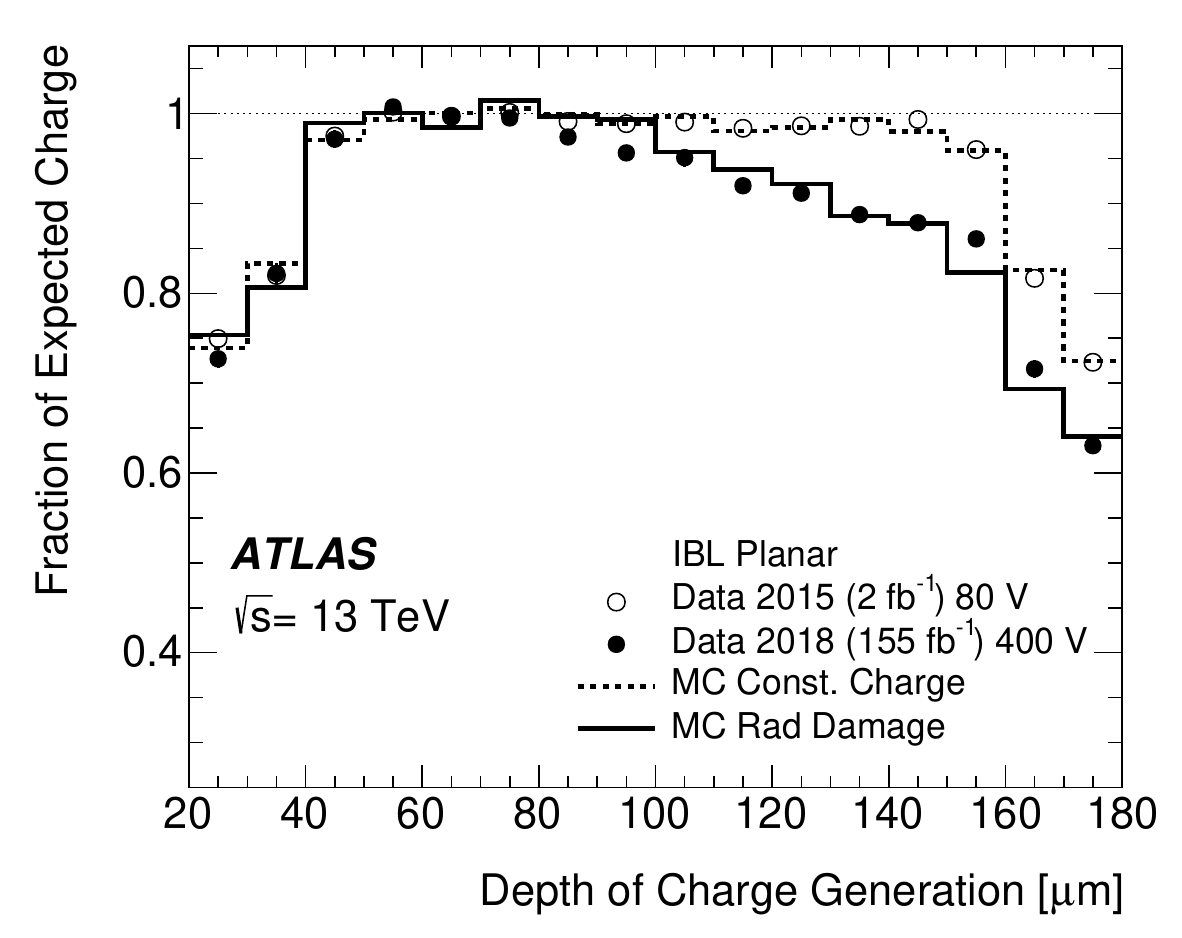}
\label{fig:Planar_Depth_Charge}
}
\caption{\protect\subref{fig:Planar_Cluster_Charge} Distribution of cluster charges in IBL planar sensors from data taken in 2022, after an estimated average fluence of about $8.7\times10^{14}$~n-eq~cm$^{-2}$, at a bias voltage of 450~V (points with error bars). The predictions from simulated events with (without) radiation damage effects are shown by the continuous (dashed) histograms. \protect\subref{fig:Planar_Depth_Charge} Charge collection efficiency as a function of the estimated depth of charge generation for IBL planar sensors in 13~TeV collision data, at the start of \RunTwo in 2015 (2~fb$^{-1}$ of integrated luminosity) shown by the open points and at the end of \RunTwo in 2018 (156~fb$^{-1}$ of integrated luminosity) shown by the filled points, and radiation damage simulation for matching fluence conditions. The efficiency is computed by comparing the fraction of the cluster charge deposited on the pixels ordered from the extrapolated point of track entrance in the Si substrate to that of track exit in the longitudinal projection. The drops observed at the two ends of the depth range are due to resolution effects. To optimise the resolution, only tracks with $\pT > 3$~GeV and at least three pixels in the longitudinal projection are considered. \label{fig:Planar_Charge}}
\end{figure}

An extensive validation of the radiation damage digitiser was performed, where the predictions for IBL planar sensors are compared with collision data. Three examples of these comparisons are shown in Figure~\ref{fig:Planar_Charge} and in Figure~\ref{fig:IBLCCE} (in Section~\ref{sec:perf-chargecoll}), where the measured cluster charge in IBL planar sensors, the charge collection efficiency (CCE) as a function of the depth of charge generation (see Section~\ref{sec:perf-chargedepth} for more details) and the evolution of the charge collection efficiency with integrated luminosity are shown in comparison with the simulation predictions.

The simulation predictions agree with the measurements using collision data over almost two orders of magnitude of radiation fluence.  The pixel radiation damage simulation is currently used in the production of ATLAS MC samples for physics analyses and it is used to predict the expected CCE as a function of integrated luminosity, to optimise operation conditions, and to calibrate the reconstruction tools.

\subsection{Simulation of radiation damage effects in IBL 3D sensors}

\begin{figure}[!htb]
\centering
\includegraphics[width=0.75\textwidth]{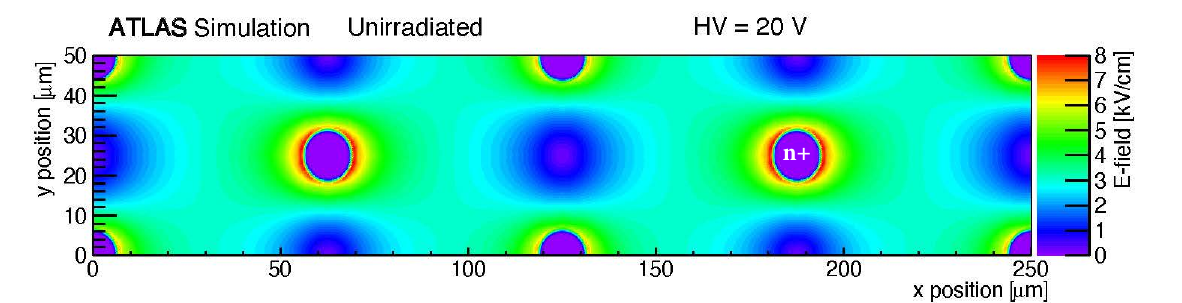}
\includegraphics[width=0.75\textwidth]{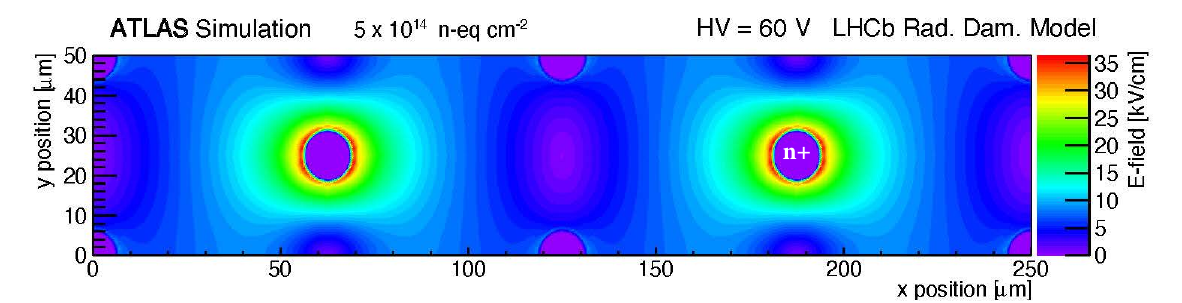}

\caption{Simulated strength of the electric field as a function of transverse (local X) and longitudinal (local Y) positions in an ATLAS IBL 3D sensor before irradiation with a bias voltage of 20~V (top panel) and for a fluence of $5\times10^{14}$~n-eq~cm$^{-2}$ with a bias voltage of 60~V (bottom panel). \label{fig:3DFieldPotential}}
\end{figure}

Radiation damage effects in the Si bulk of 3D pixel sensors are simulated using a procedure similar to that used for the planar sensors and discussed above. Since the p-type material, used in 3D sensors, does not undergo type inversion due to irradiation, unlike the n-type material used in planar sensors, a different TCAD radiation damage model is needed. The LHCb radiation damage model~\cite{FOLKESTAD201794} is used. In contrast to planar sensors, the field in 3D sensors of FBK design is nearly independent of the depth and depends strongly on the position on the sensor surface. The n$^+$ and p$^+$ columns are characterised by a negligible field magnitude and they are modelled as completely inefficient volumes. The computation of the Ramo potential for 3D sensors is also  more complex than that for planar sensors because the calculation requires a relatively large simulation area. As in the case of planar sensors, only the immediate neighbours are included in the calculation. An example of the electric field simulated for an unirradiated 3D pixel and an irradiated 3D pixel is shown in Figure~\ref{fig:3DFieldPotential}.

The time taken by charge carriers to reach their collection implant from the ionisation point is computed from these field maps. Compared to planar sensors, electrons and holes follow non-trivial trajectories in 3D devices, due to the more complex shape of the electric field.  However, a simplification is afforded in the simulation, since the electric field lines run nearly parallel to the solenoidal magnetic field and the Lorentz angle is negligibly small.

\begin{figure}[!hb]
\centering
\includegraphics[width=0.65\textwidth]{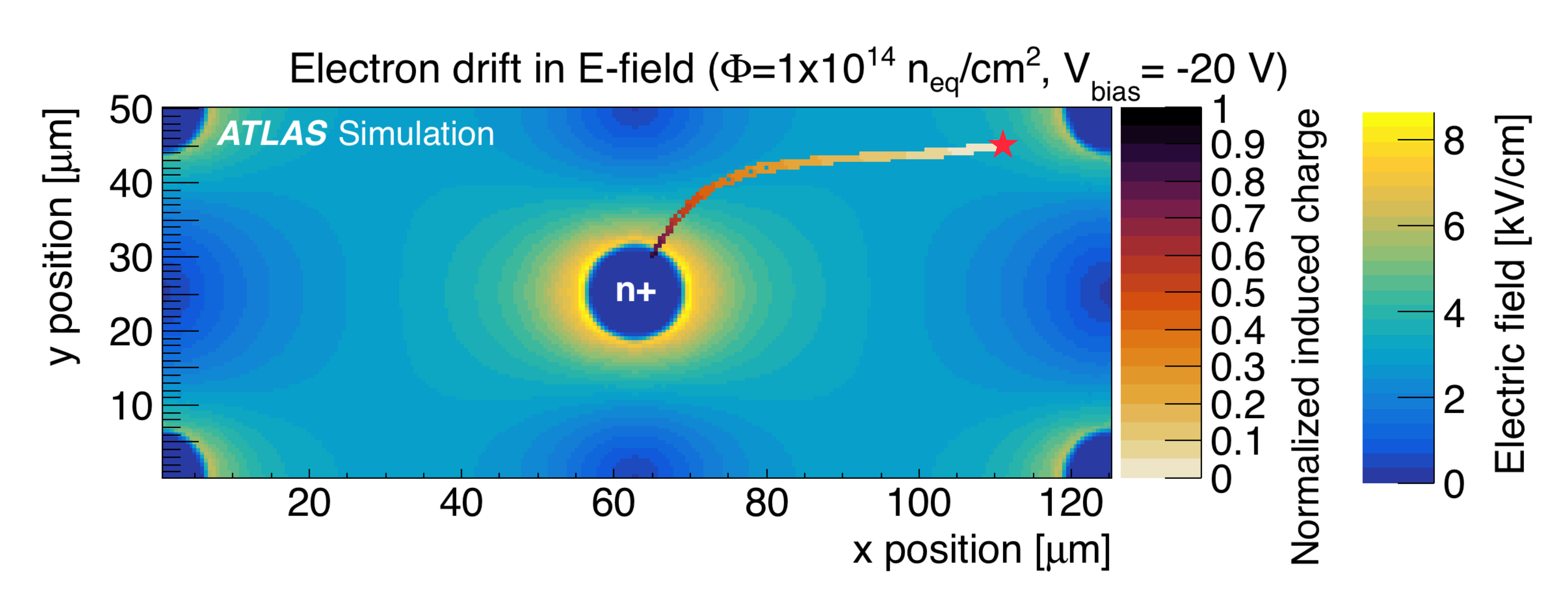}
\caption{Electric field and normalised induced charge in one-half of a 3D sensor for a simulated bias voltage of $-20$~V.  The initial electrons are ionised in the top right corner of the plot, indicated by a star.  Under the influence of the electric field, they drift toward the n$^+$ electrode in the centre. During this drift, the electrons may get trapped. The markers indicate the location of trapped charges and the marker colour shows the corresponding fraction of the deposited charge induced on the read-out electrode. The process is repeated multiple times, accounting for diffusion effects. From Ref.~\cite{IDET-2017-10}.\label{fig:induced_signal.pdf}}
\end{figure}

The combined effect of the electric field, charge trapping, and Ramo potential is illustrated in Figure~\ref{fig:induced_signal.pdf} from Ref.~\cite{IDET-2017-10}, where the signal induced by electrons drifting towards the n$^+$ electrode is shown in terms of the normalised induced charge, defined as the fraction of the deposited charge that is induced on the read-out electrode. The simulation is repeated multiple times by taking the same initial position of the electrons to show the effect of trapping on the induced signal. The closer to the collecting electrode the final carrier position is, the larger the induced signal. From this figure it is also possible to appreciate the complex path that the collected electrons follow, as a result of the shape of the electric field.

CNM sensors have columns that do not fully traverse the Si bulk, and the electric field and Ramo potential are not invariant with translation in Si sensor depth. This feature is currently not implemented in the radiation damage simulation model. Therefore, the comparisons of simulation predictions with collision data are performed for FBK sensors with fully traversing columnar electrodes.


\section{3D Sensor Performance and Radiation Damage Effects}
\label{sec:perf}


\subsection{Datasets and analysis procedure}
\label{sec:perf-datasets}

The study of the IBL 3D sensor performance is based on four datasets: the $\sqrt{s} = 13$~TeV $pp$ collision data collected during \RunTwo from 2015 to 2018, the cosmic-ray data collected in milestone runs during the LHC LS2 from 2019 to 2022, and the $\sqrt{s} = 900$~GeV and 13.6~TeV $pp$ collision data collected during \RunThr operations in 2022 and 2023. These data offer the opportunity to investigate the 3D sensor response under different conditions and an increasing amount of delivered fluence. Since the 3D sensors cover a limited acceptance region at large $|\eta|$ values, their study in collision data is affected by the smaller event sample available and the small incidence angles of tracks traversing their active volume. This makes cosmic-ray and 900~GeV data, characterised by larger rates of tracks traversing the 3D sensors at larger incidence angles, valuable additions for this study. Due to the uniform illumination of the IBL detector by cosmic-ray muons, approximately 22\% of IBL hits associated with a selected track are registered on a 3D sensor. Collisions at 900~GeV also yielded a significant fraction of tracks traversing the 3D pixels, 1.5\%, due to the long beam-collision region in $z$. These fractions should be compared with the 0.7\% of selected particle tracks with an IBL hit in the 3D sensors in high-energy collision events.

The software infrastructure developed for the reconstruction process of real detector data and MC simulation are presented in Ref.~\cite{ATLAS:2024vdo}. In the reconstruction, pixel clusters are built from adjacent pixels reporting in-time charge above a preset analogue threshold. This threshold was set at 2.5~k$e^-$ and then 2.0~k$e^-$ during \RunTwo and at 1.5~k$e^-$ during LS2 and in \RunThr operations. In this study, only pixel clusters reconstructed in the 3D modules operating at the nominal bias voltage and associated with a reconstructed charged-particle track (\enquote{hits on track}) are considered. Reconstructed charged-particle tracks are accepted if they fullfil a set of quality criteria whose definition depends on the dataset under analysis. Since the cluster charge depends on the particle's track length in the active Si bulk, the charge is rescaled by $\cos \alpha$, where $\alpha$ is the particle's angle of incidence on the sensor surface.

Particle tracks from collision events are selected by requiring transverse momentum, \pT, in excess of 0.7~GeV, at least two hits in the Pixel layers, with at least one in the IBL, and a total of at least seven hits in the Pixel layers and the Si strip layers in the SCT. In addition, pixel clusters shared or split between two or more tracks are not considered for this study.

Samples of $Z$\,+\,jets, $Z \rightarrow \mu\mu$ and dijet QCD events simulated at 13 and 13.6~TeV were generated with \Pythia[8]~\cite{Bierlich:2022pfr} and reconstructed after applying the ATLAS detector simulation. This simulation included the pixel radiation damage digitiser, discussed in Section~\ref{sec:raddamage}, for the signal formation. Reconstructed charged-particle tracks in simulation are selected using the same criteria as used for data. In addition, simulated particle tracks are reweighted to reproduce the distribution of the incidence angle $\alpha$ on the IBL 3D sensor surface measured in data.

Cosmic-ray muons provide a large sample of high-momentum particle tracks and have played an important role in the commissioning of the detector~\cite{Cosmics2010}. More recently, they were important for understanding the detector performance during shutdown periods. During cosmic-ray data-taking periods of LS2, from 2020 to 2022, regularly performed bias-voltage scans of the pixel detectors extended the voltage range for 3D pixels to values higher than those used when collecting 13~TeV collision data. Muon tracks are reconstructed as follows. First, particle tracks are reconstructed starting with the hits in the Pixel and SCT detectors. These silicon-only tracks are then extrapolated to the Transition Radiation Tracker (TRT) and re-fitted using all associated space-points. A special pattern-recognition algorithm is used to reconstruct cosmic-ray muons as single tracks on both sides of the central beam-axis region with no cut placed on the track's transverse impact parameter, $d_0$. Only events with one track reconstructed in the ATLAS Inner Detector are selected. Cosmic-ray muon tracks are then split into two halves by fitting two new tracks, each containing only the hits on one side of the beam axis (\enquote{split tracks}). Split muon tracks are required to have at least two hits in the Pixel layers, with at least one in the IBL, and a total of at least five hits in the Pixel and SCT layers. Split tracks are advantageous for performance studies since they have a hit content comparable to that of tracks from $pp$ collision events.

In the following, the integrated luminosity values refer to the integrated luminosity delivered by the LHC since the time of IBL installation. As discussed in the previous section, this integrated luminosity can be converted to the neutron-equivalent fluence on the 3D sensors.

\subsection{Charge collection and pixel multiplicity in clusters}
\label{sec:perf-chargecoll}

\begin{figure}[htb]
\centering
\subfloat[]{
\hspace*{-0.55cm} \includegraphics[width=0.34\textwidth]{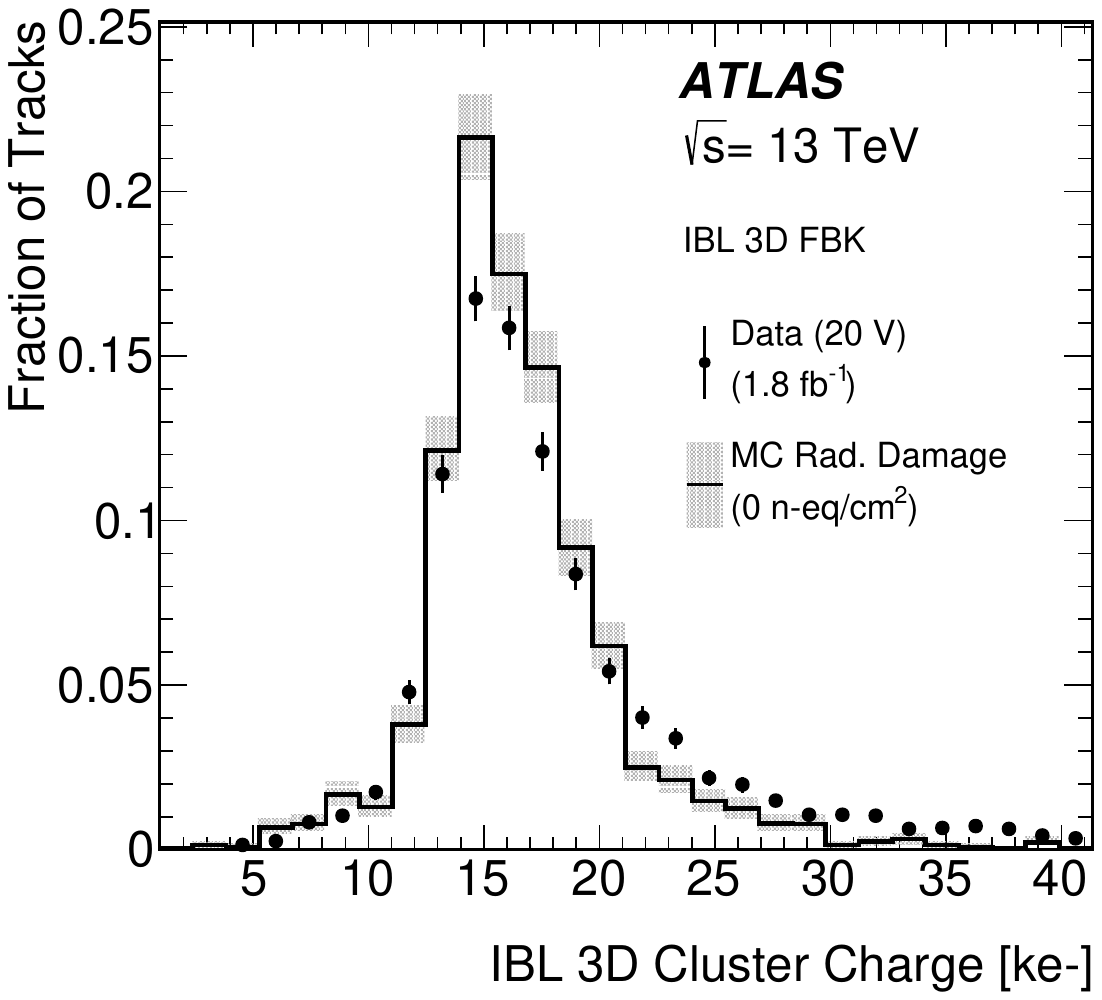}
\label{fig:3DCharge2015}
}
\subfloat[]{
\hspace*{-0.55cm} \includegraphics[width=0.34\textwidth]{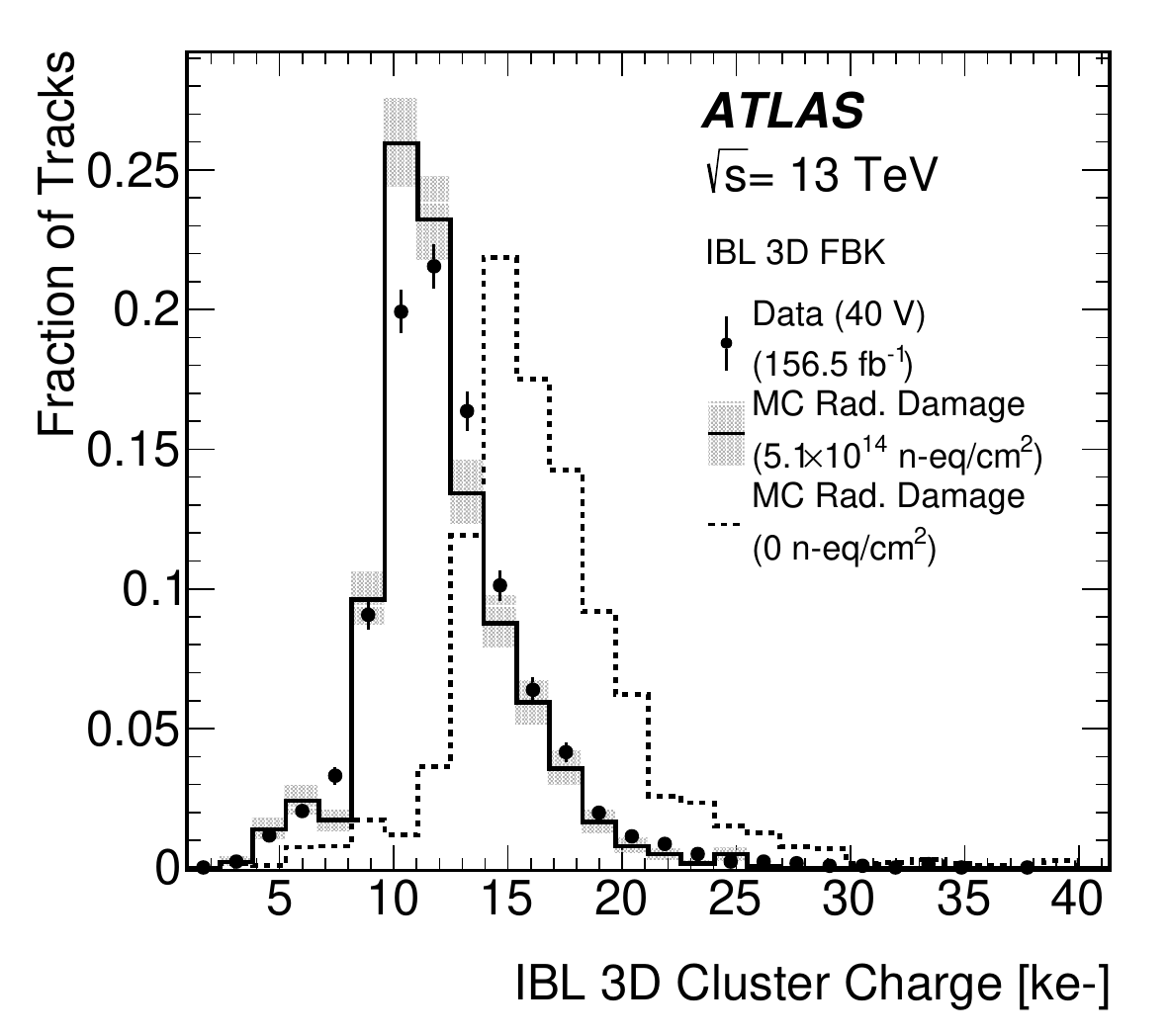}
\label{fig:3DCharge2018}
}
\subfloat[]{
\hspace*{-0.55cm} \includegraphics[width=0.34\textwidth]{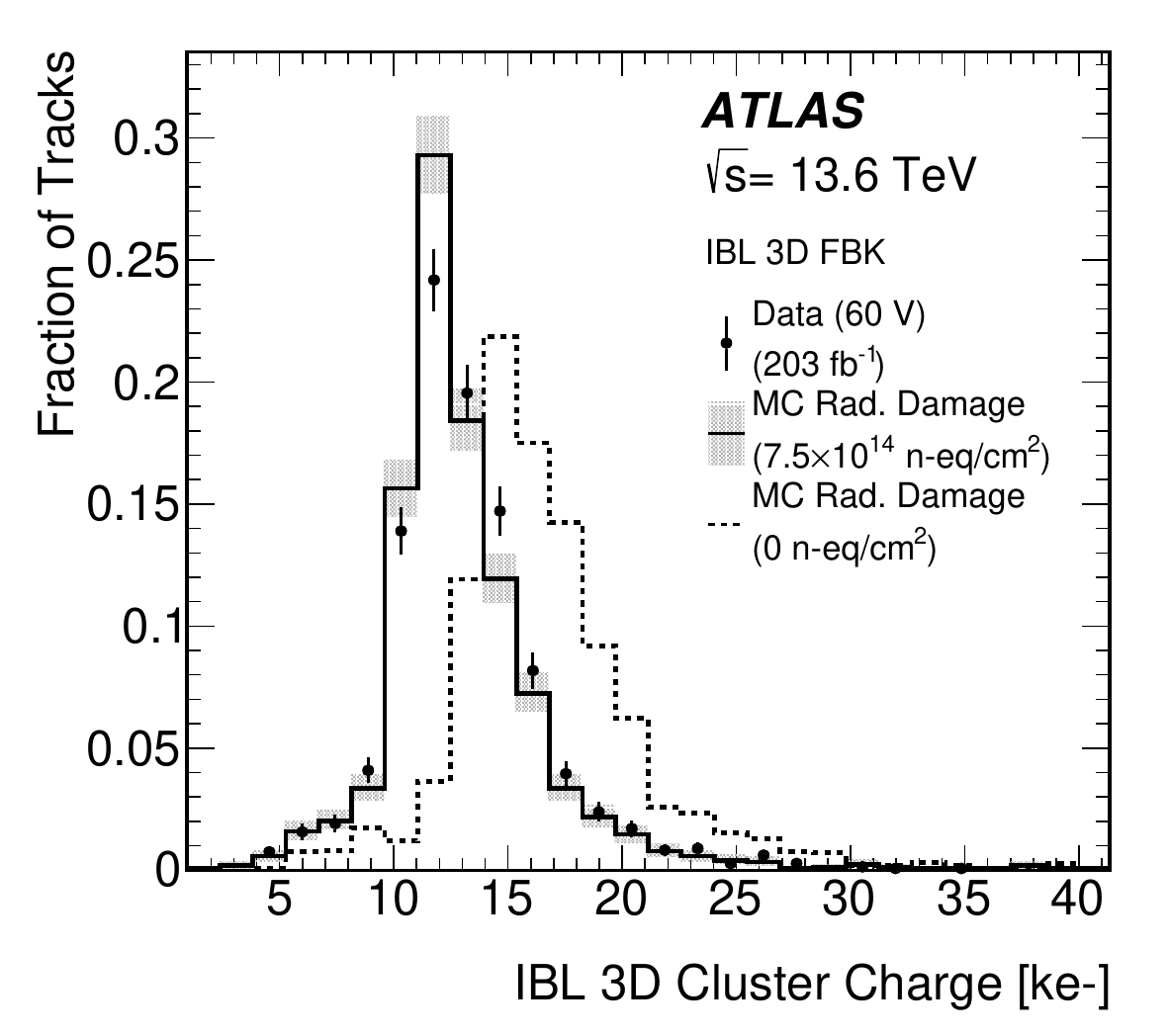}
\label{fig:3DCharge2023}
}

\caption{Cluster charge corrected for particle path length in the Si for 3D FBK sensors in $pp$ collision events. Points with error bars represent data recorded after \protect\subref{fig:3DCharge2015} 1.8~fb$^{-1}$ of integrated luminosity with 20~V bias voltage in 2015, \protect\subref{fig:3DCharge2018} 156.5~fb$^{-1}$ with 40~V in 2018 and \protect\subref{fig:3DCharge2023} 203~fb$^{-1}$ with 60~V in 2023, compared with the predictions of ATLAS radiation damage simulation for fluences of \protect\subref{fig:3DCharge2015} 0~n-eq~cm$^{-2}$, \protect\subref{fig:3DCharge2018} 0 and $5.1 \times 10^{14}$~n-eq~cm$^{-2}$, and \protect\subref{fig:3DCharge2023} 0 and $7.5 \times 10^{14}$~n-eq~cm$^{-2}$, shown by the histograms.~\label{fig:3DCharge}}
\end{figure}

The charge distribution for clusters in IBL 3D sensors associated with reconstructed particle tracks is shown in Figure~\ref{fig:3DCharge} for tracks in collision data recorded at 13~TeV in 2015 and 2018, and at 13.6~TeV in 2023, compared with predictions from radiation damage simulations for the corresponding values of particle fluence and also with the prediction in the absence of radiation damage effects. The shape of the collected charge distribution depends on several factors, such as threshold dispersion, particle incidence angles, particle-species composition and charge calibration, that may differ between data and MC simulation but are not linked to radiation damage effects.  The overall modelling of the data after radiation damage, and the decrease in charge collection efficiency, is quite good.

\begin{figure}[hb!]
\centering
\subfloat[]{
\includegraphics[width=0.475\textwidth]{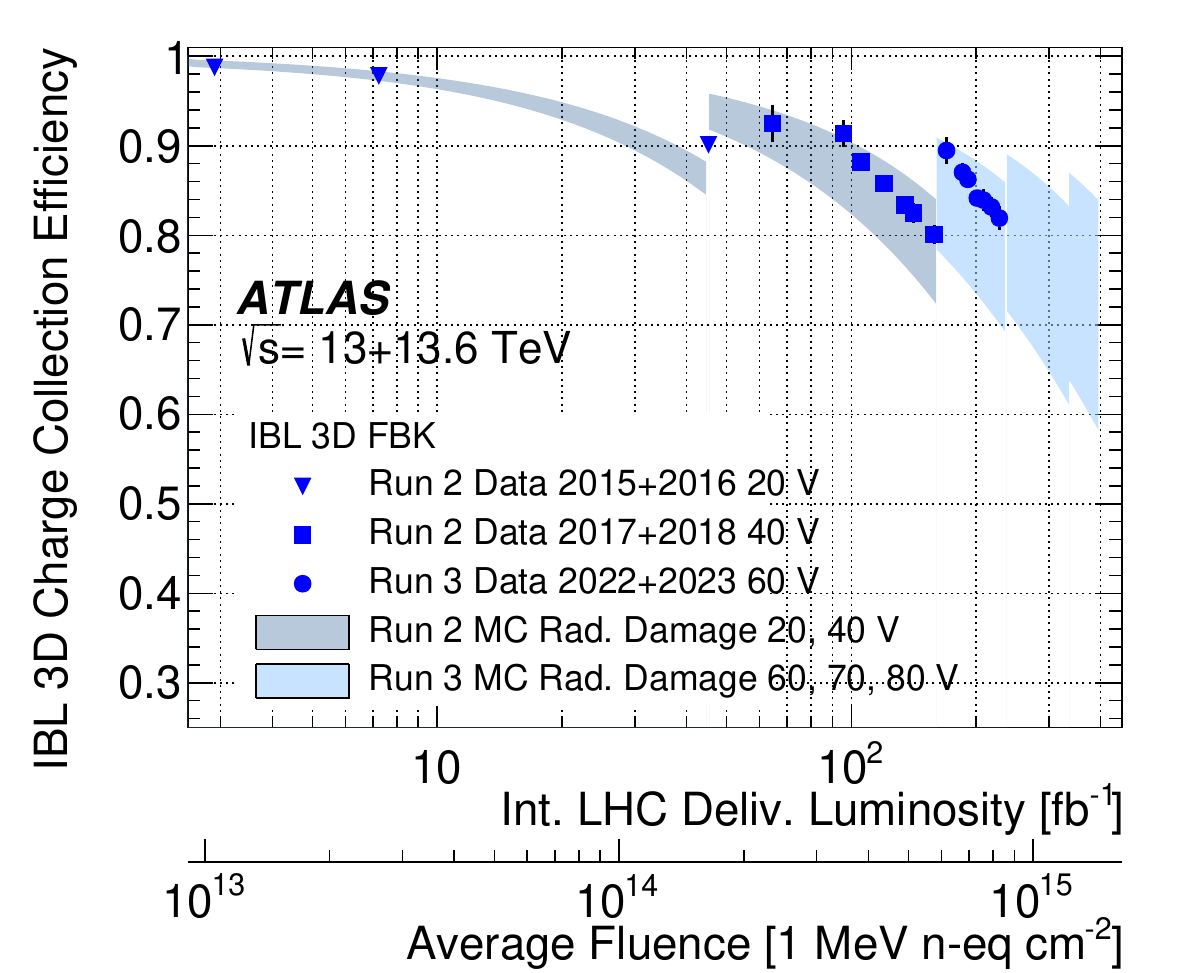}
\label{fig:IBL3DCCE}
}
\subfloat[]{
\includegraphics[width=0.475\textwidth]{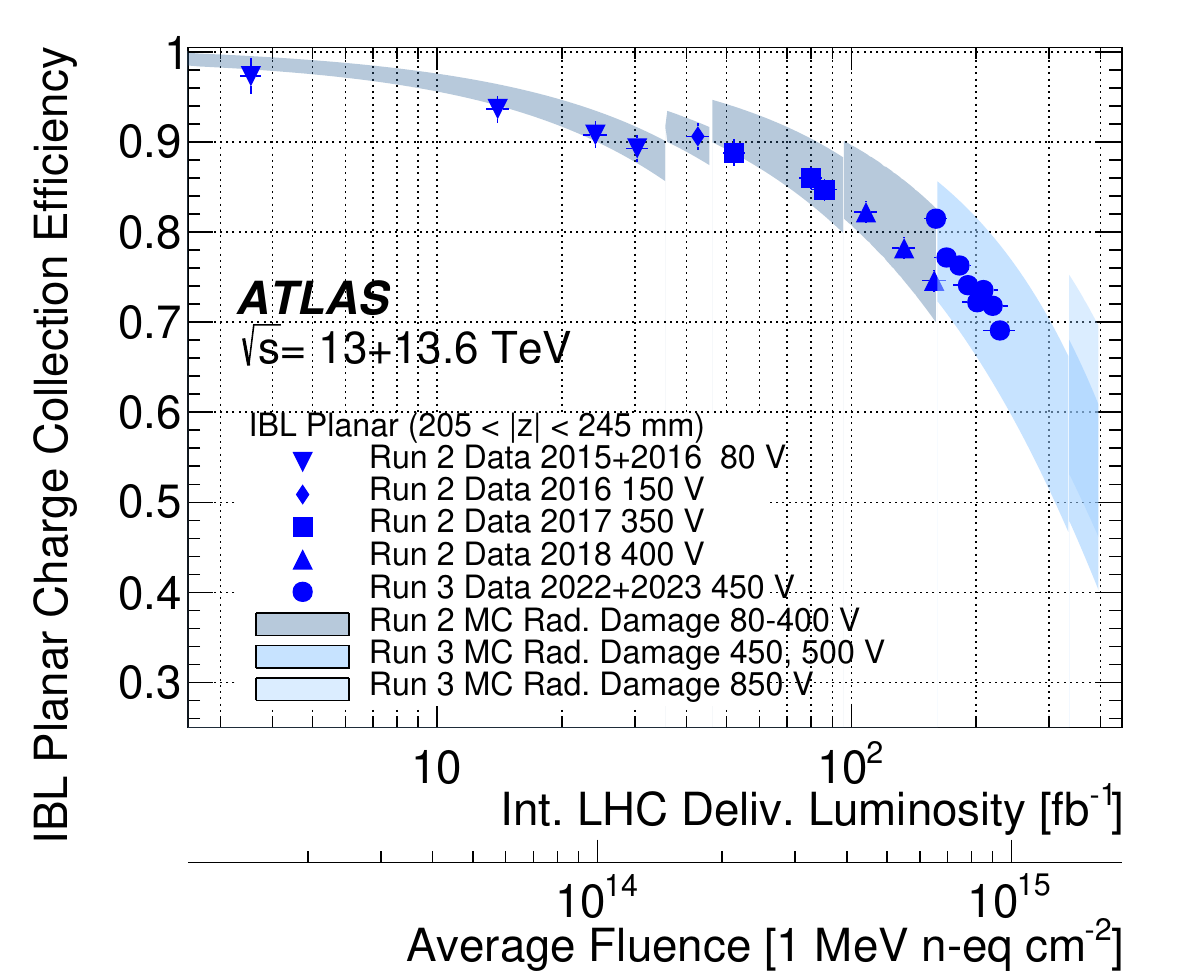}
\label{fig:IBLPLM6CCE}
}
\caption{Charge collection efficiency as a function of the delivered integrated luminosity and average particle fluence for \protect\subref{fig:IBL3DCCE} IBL 3D sensors of FBK design and \protect\subref{fig:IBLPLM6CCE} IBL planar sensors, installed next to the 3D sensors at 205\,$<|z|<$\,245~\si{mm}, for data and ATLAS radiation damage simulation from the beginning of \RunTwo. The points represent the data and the bands show the simulation's predictions with their uncertainties. The fluence is extracted from leakage current data and averaged over the longitudinal position along the detector modules using the same distribution of longitudinal positions of impact as the particle tracks used in the analysis. Sudden increases in charge collection efficiency at the beginning of each year are due to changes in the operational parameters (bias voltage and thresholds). Predictions for the evolution in 2024 and 2025 are also given. \label{fig:IBLCCE}}
\end{figure}

The most probable value (MPV) of the charge collected in the clusters is a measure of the amount of collected charge and its agreement between data and radiation damage simulation validates MC predictions of charge collection efficiency with increasing fluence. The MPV is extracted from the cluster charge distributions by performing a fit with a Landau function, representing the collected signal charge distribution, folded with a Gaussian function centred on zero and of variable width, representing the charge resolution effects. These effects are due to thresholds, charge calibration and electronic noise. The fit has four free parameters: the Landau peak position and width, the Gaussian noise width and a normalisation term. The fitted Landau peak position is the estimator of the collected signal charge MPV.

These values are used to measure the charge collection efficiency, defined as the ratio of the MPV of the charge collected at a given point in time to that measured for undamaged sensors at the start of \RunTwo, when the MPV for FBK sensors was measured to be $(14.59\pm0.06)$~k$e^-$ in data and $(14.71\pm0.12)$~k$e^-$ in simulation (see Figure~\ref{fig:3DCharge2015}). Its evolution with the integrated luminosity and fluence was studied in data and simulation. The same analysis was repeated by selecting the IBL modules equipped with planar sensors installed closest to the 3D modules, covering the longitudinal region, along the beam axis, of 205\,$<|z|<$\,245~\si{mm}. These modules receive an estimated track-averaged fluence of $(4.1 \pm 0.4) \times 10^{12}$~n-eq~cm$^{-2}$/fb$^{-1}$, close to the value of $(3.6 \pm 0.4) \times 10^{12}$~n-eq~cm$^{-2}$/fb$^{-1}$ estimated for the 3D sensors, making the performance comparison between the two technologies less dependent on the assumed longitudinal fluence profile. The results are summarised in Figure~\ref{fig:IBLCCE}. The uncertainty band shown for the simulation's prediction includes the effects of parametric uncertainties due to the trapping constant, the conversion of integrated luminosity to fluence, and the estimate of the electric field in the Si bulk. The measured evolution and predicted evolution of the charge collection efficiency with integrated luminosity and fluence for 3D and planar sensors are in good overall agreement for fluences spanning two orders of magnitude.

\begin{figure}[h!]
\centering
\subfloat[]{
\includegraphics[width=0.485\textwidth]{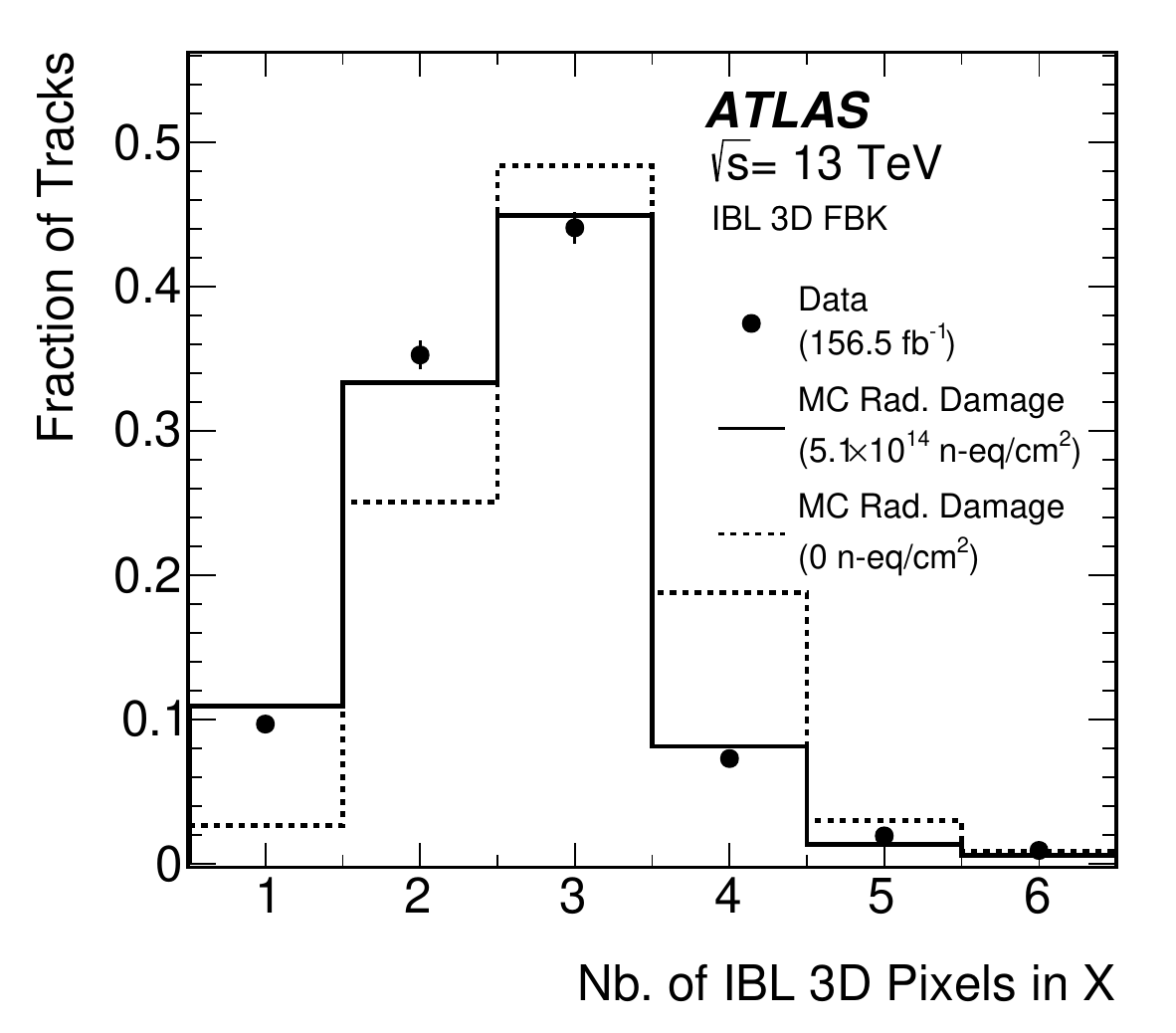}
\label{fig:3DNPixX}
}
\subfloat[]{
\includegraphics[width=0.485\textwidth]{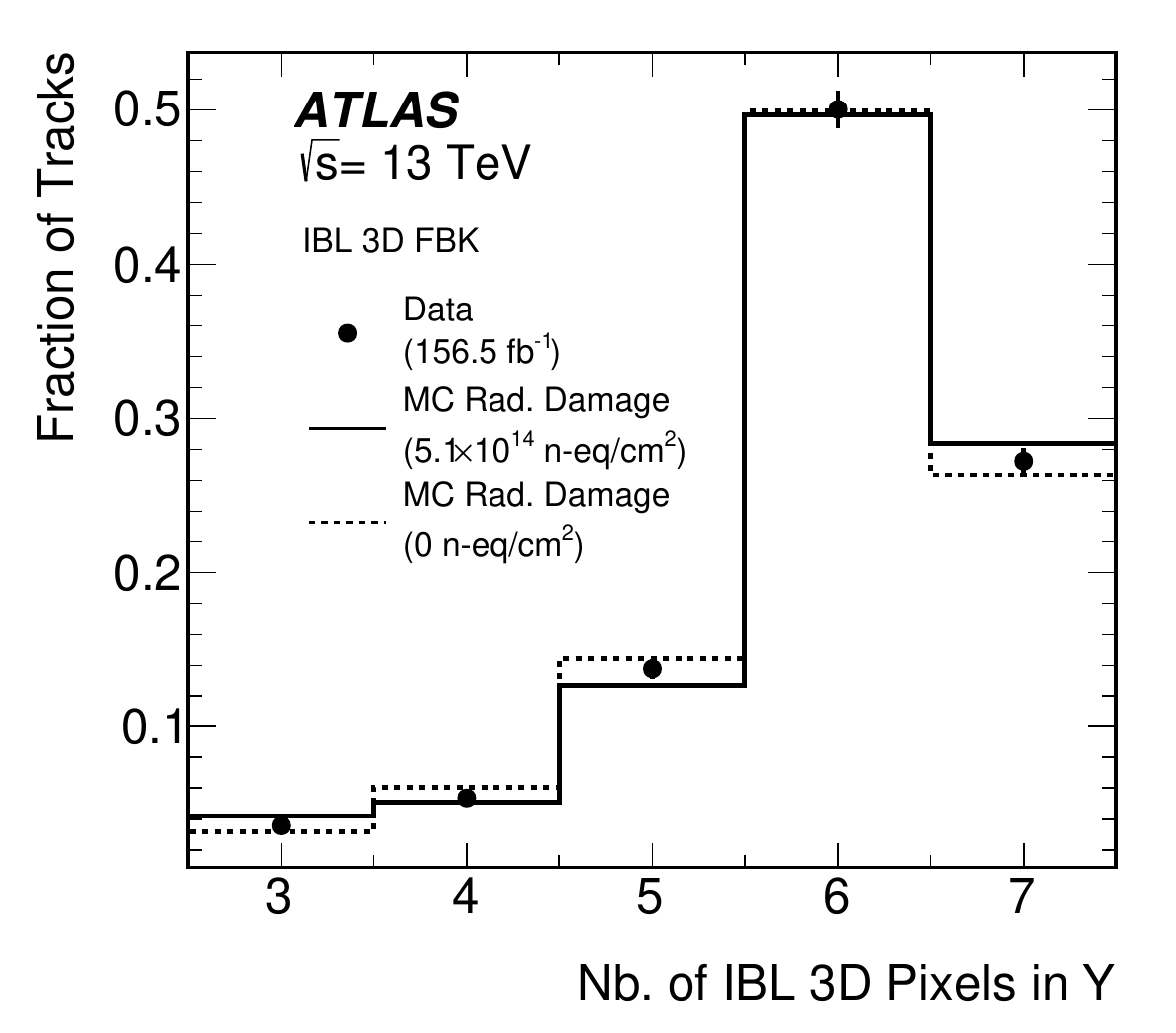}
\label{fig:3DNPixY}
}
\caption{Number of pixels in the \protect\subref{fig:3DNPixX} transverse (local X) and  \protect\subref{fig:3DNPixY} longitudinal (local Y) projection for IBL 3D sensors with clusters associated with reconstructed particle tracks. Points with error bars represent data in 13~TeV collision events collected in 2018 after 156.5~fb$^{-1}$ of integrated luminosity and are compared with the predictions of ATLAS radiation damage simulation for fluences of 0 (dashed line) and $5.1 \times 10^{14}$~n-eq~cm$^{-2}$ (continuous line). The larger pixel multiplicity in the longitudinal projection is dominated by the affect of the shallow angle of incidence of particle tracks in the IBL module region equipped with 3D sensors.~\label{fig:3DNPix}}
\end{figure}

These results are of special interest since the 3D response at fluences in the $10^{14}$~n-eq~cm$^{-2}$ range, relevant to the study of \RunTwo and early \RunThr data, was not studied with irradiation and beam-test campaigns on FE-I4 assemblies in the R\&D phase. Instead, results for these fluences are available for sensors fabricated for use in the CMS experiment. One of the geometrical configurations with two electrodes per pixel, in the so-called \enquote{2E layout}, has an inter-electrode distance of 62.5~$\mu$m, similar to that of the ATLAS IBL configuration (67.3~$\mu$m), and can be used for a valid comparison with the ATLAS IBL \RunTwo data. These sensors were irradiated up to a fluence of $7\times10^{14}$~n-eq~cm$^{-2}$. At this fluence the charge collection efficiency was observed to drop to $\simeq$\,80\% of its pre-irradiation value at a bias voltage of $\sim$\,35~V~\cite{Bubna:2014ria,Obertino:2013tia,Bubna:201352}. These results are close to the CCE values measured for the IBL 3D pixels at the end of \RunTwo and those predicted by radiation damage simulation.

The study of charge collection efficiency using collision data and radiation damage simulation provides a direct comparison of how this efficiency evolves with fluence on pixels in 3D and planar technologies. The IBL 3D pixels show better response in terms of charge collection efficiency for the same fluence, with CCE ratios to planar sensors of $1.13\pm0.03$ and $1.15\pm0.03$ for pixels of FBK and CNM design, respectively, at $8.3\times10^{14}$~n-eq~cm$^{-2}$. The uncertainties in these measurements include the contribution from the systematic uncertainty in the integrated luminosity-to-fluence conversion factors. For comparison, the radiation damage simulation predicts a ratio of $1.10\pm0.03$ at $8.3\times10^{14}$~n-eq~cm$^{-2}$. This is predicted to increase to $1.27\pm0.06$ at the fluence of $1.5\times10^{15}$~n-eq~cm$^{-2}$ expected on the IBL 3D sensors at the end of \RunThr, where the quoted value assumes the 3D pixels operate at 80~V in 2025 and its uncertainty includes the effect of a $\pm$10~V bias voltage change.

The 3D sensors are expected to have less charge spread than planar pixel sensors, due to the specific electric field distribution in the bulk.
\begin{figure}[ht!]
\centering
\subfloat[]{
\includegraphics[width=0.45\textwidth]{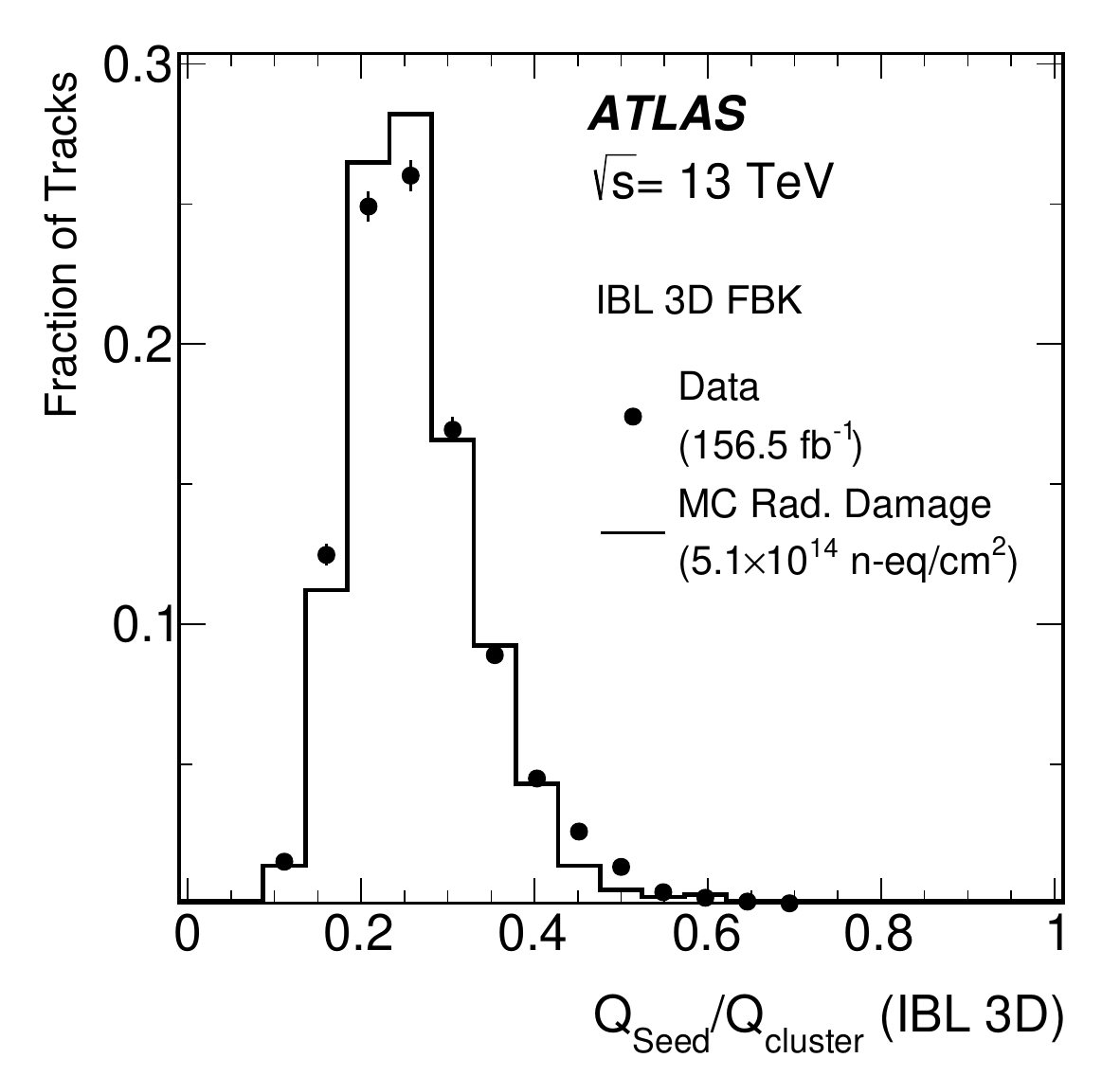}
\label{fig:3DChargeSeedFrac}
}
\subfloat[]{
\includegraphics[width=0.45\textwidth]{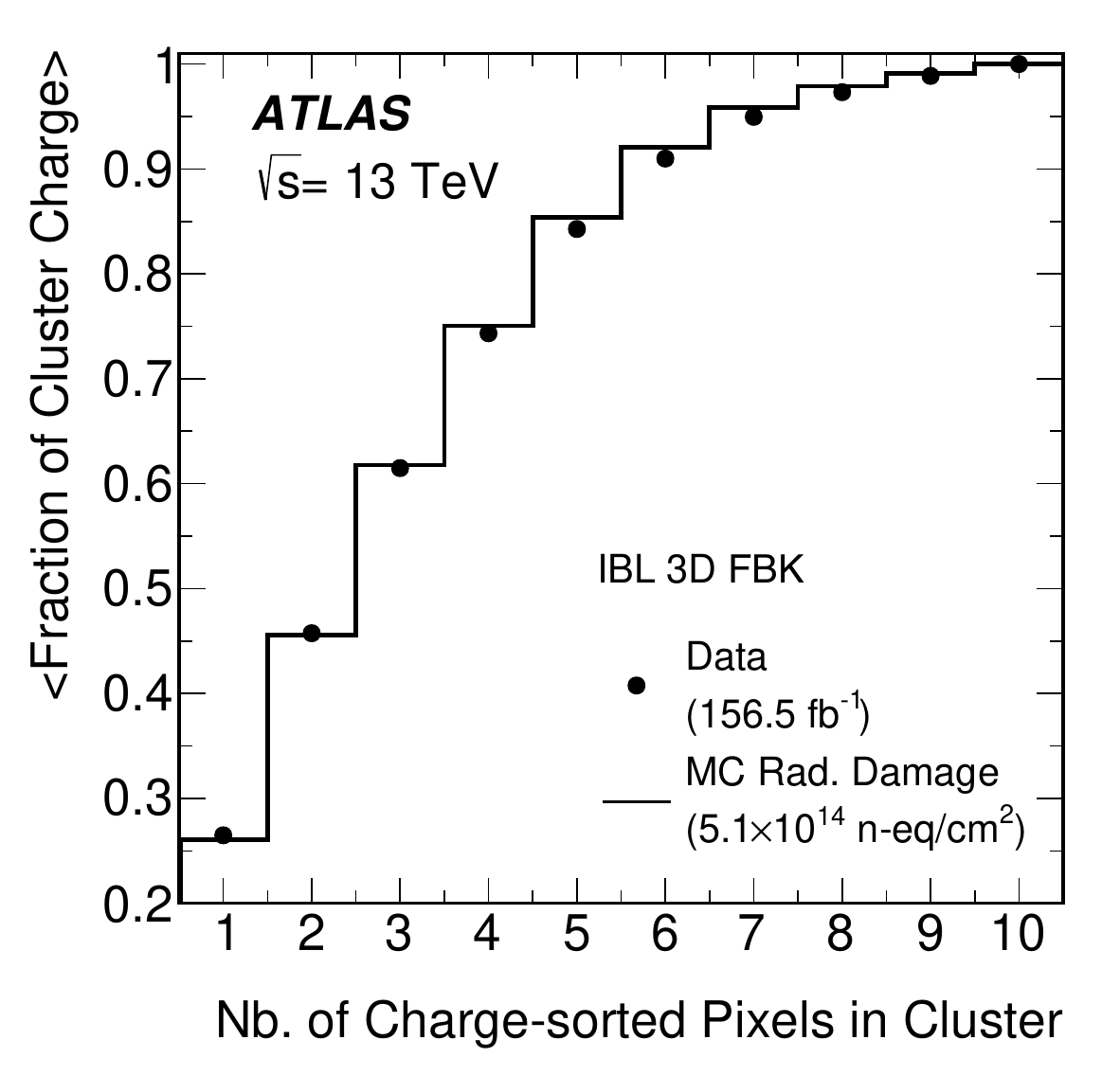}
\label{fig:3DChargeIntFrac}
}
\caption{Charge sharing in 3D clusters for 3D FBK sensors in 13~TeV collision events collected after 156~fb$^{-1}$ of integrated luminosity: \protect\subref{fig:3DChargeSeedFrac} fraction of the total cluster charge carried by the pixel with the highest charge (\enquote{seed pixel}) and \protect\subref{fig:3DChargeIntFrac} average cumulative fraction of the cluster charge as a function of the number of contributing pixels, sorted in order of decreasing pixel charge. Points represent data and the histogram shows the radiation damage simulation.~\label{fig:3DChargeFrac}}
\end{figure}
The distributions of the number of pixels in the cluster in the transverse and longitudinal projections for data and simulation are shown in Figure~\ref{fig:3DNPix}. The charge distribution in clusters is studied using two related observables. The first is the fraction of the total cluster charge carried by the leading pixel in the cluster. The distribution of this fraction for selected 3D pixel clusters in collision events is compared for data and simulation in Figure~\ref{fig:3DChargeSeedFrac}. The second observable sensitive to the charge spread in clusters is the average cumulative fraction of the cluster charge as a function of the number of contributing pixels, sorted in order of decreasing pixel charge. This distribution is sensitive to charge sharing to pixels beyond the leading one, and represents the average charge spread over the full cluster (Figure~\ref{fig:3DChargeIntFrac}).

\subsubsection{Charge collection vs.\ bias voltage}
\label{sec:perf-chargehv}

The depletion voltage for unirradiated IBL 3D sensors is expected to be in the range $\simeq$\,5--10~V. In 2015, at the beginning of \RunTwo, they were overdepleted with a bias voltage of 20~V. As the Si bulk suffered radiation-induced damage, the depletion voltage increased with the radiation fluence. Changes in the charge collected in the reconstructed clusters with applied bias voltage were studied during \RunTwo, LS2 and \RunThr.
\begin{figure}[ht!]
\centering
\includegraphics[width=0.55\textwidth]{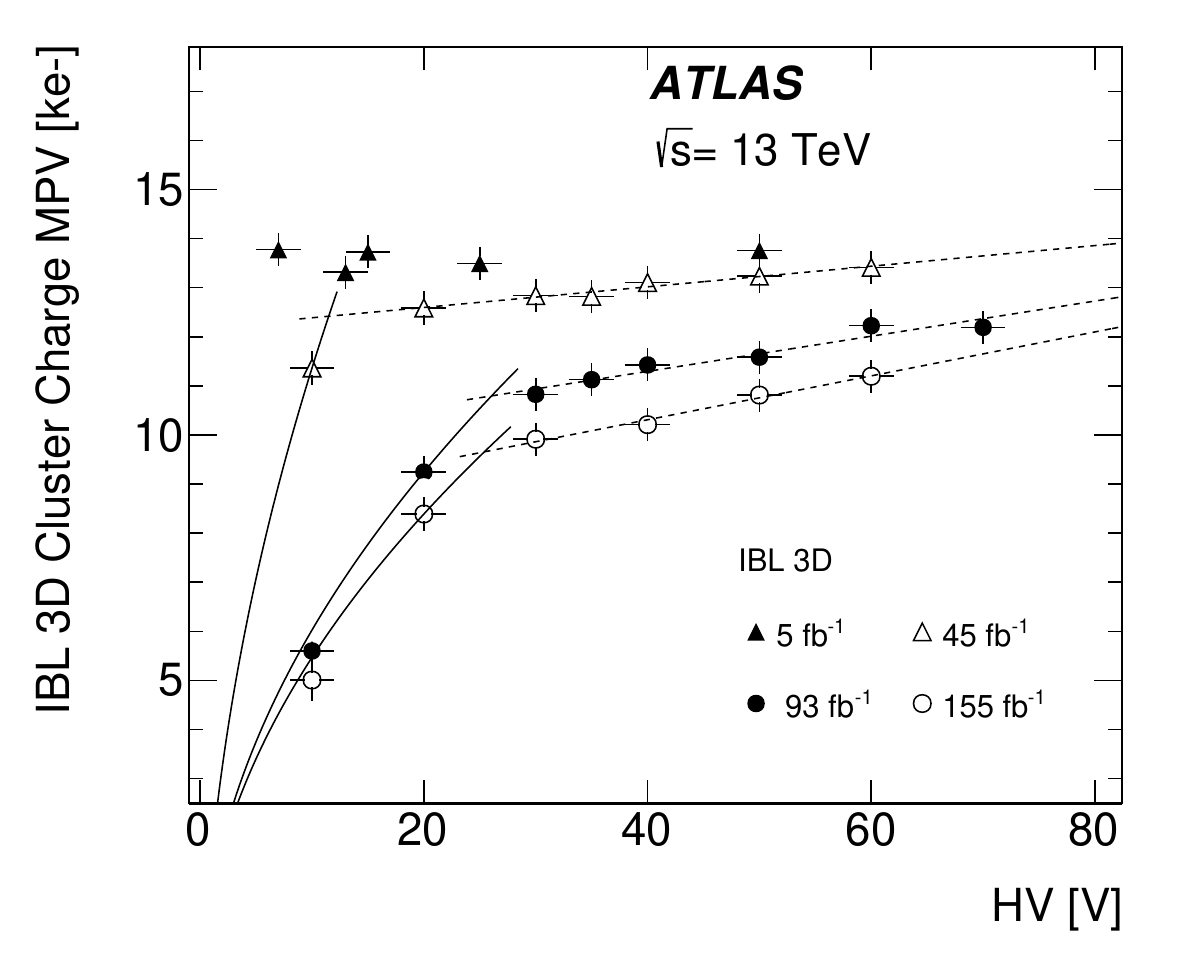}
\caption{Most probable value (MPV) of the pixel cluster's charge, corrected for particle path length in the Si, for IBL 3D sensors with clusters associated with reconstructed particle tracks as a function of the applied detector bias voltage (HV). Points represent data recorded during HV scans during \RunTwo, after delivered integrated luminosities ranging from 5~fb$^{-1}$, in early 2016, to 155~fb$^{-1}$, towards the end of 2018, corresponding to average particle fluences of $(0.2{-}5)\times 10^{14}$~n-eq~cm$^{-2}$ on the sensors. The \enquote{depletion voltage} is defined by the HV value where the MPV values transition from square root (continuous lines) to linear (dashed lines) behaviour.~\label{fig:3DChargeVsHV}}
\end{figure}
Collision events were collected at various bias voltages during voltage scans performed in \RunTwo at the beginning of the 2016 run, with only 5~fb$^{-1}$ of integrated luminosity, and at the end of the 2016, 2017 and 2018 LHC $pp$ collision periods, corresponding to 45, 93 and 155~fb$^{-1}$ of integrated luminosity, and also in \RunThr at the end of 2022 and 2023, corresponding to 194 and 228~fb$^{-1}$ of integrated luminosity, respectively. During these voltage scans the pixel bias voltage, $V$, was varied in ranges from 5~V to 90~V. The evolution of the MPV of the cluster charge for 3D clusters on tracks as a function of the bias voltage in the \RunTwo datasets is shown in Figure~\ref{fig:3DChargeVsHV}.

The value of the \enquote{depletion voltage} is extracted from these data as follows. For an undepleted detector the collected charge increases $\propto \sqrt{V}$, with the thickness of the depletion region. An undamaged detector sees the collected charge increase until the Si bulk is fully depleted, corresponding to the depletion voltage value, at which point the charge saturates and becomes independent of the applied bias voltage. This saturation is observed in Figure~\ref{fig:3DChargeVsHV} for the data from the HV scan performed after just 5~fb$^{-1}$ of integrated luminosity. If a detector has significant radiation damage, the definition of depletion voltage requires some attention. In fact, in a damaged Si detector, the amount of collected charge continues to increase, roughly $\propto V$, above the point of full Si depletion due to the reduction of the charge trapping effect with the increasing charge carrier velocity. In the present study, the depletion voltage is defined as the bias voltage at which the transition between the $\propto \sqrt{V}$ and $\propto V$ regimes is observed. This is obtained by performing an iterative $\chi^2$ fit of a linear $a + b V$ function and a $c + d \sqrt{V}$ function, fitting the points from the upper and lower ends of the scan range, respectively, and repartitioning the points until the configuration of minimum $\chi^2$ is found.
The continuous and dashed lines in Figure~\ref{fig:3DChargeVsHV} represent the fitted functions modelling the increase in charge collection efficiency in the regimes below and above the depletion point, respectively. The bias voltage value corresponding to the point of intersection of the two functions is taken as the estimate of the detector's depletion voltage. With increasing fluence, the depletion voltage can be seen to increase, together with a progressive increase of the slope of the linear charge rise with voltage above full depletion.
\begin{figure}[ht!]
\centering
\subfloat[]{
\includegraphics[width=0.485\textwidth]{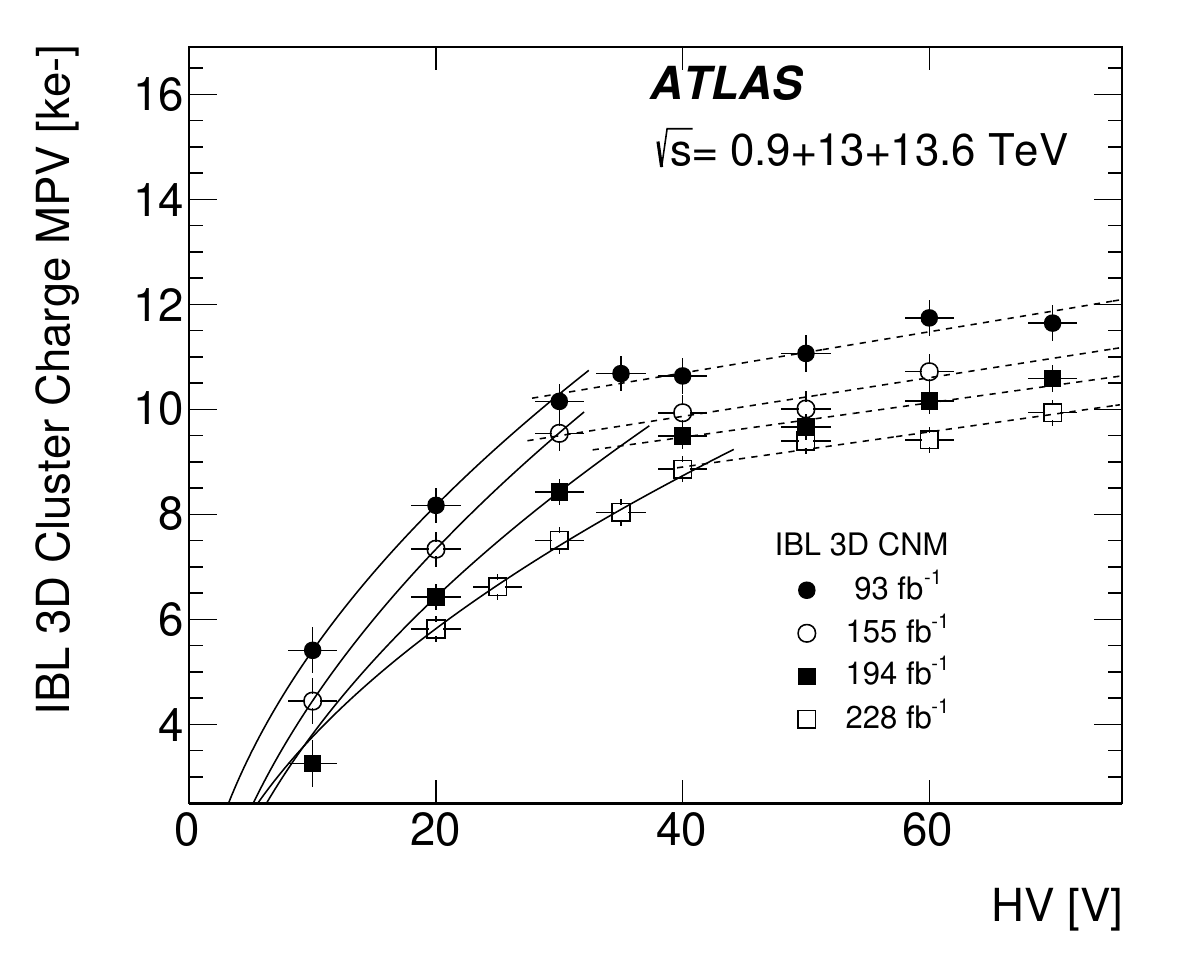}
\label{fig:3DCNMChargeVsHV}
}
\subfloat[]{
\includegraphics[width=0.485\textwidth]{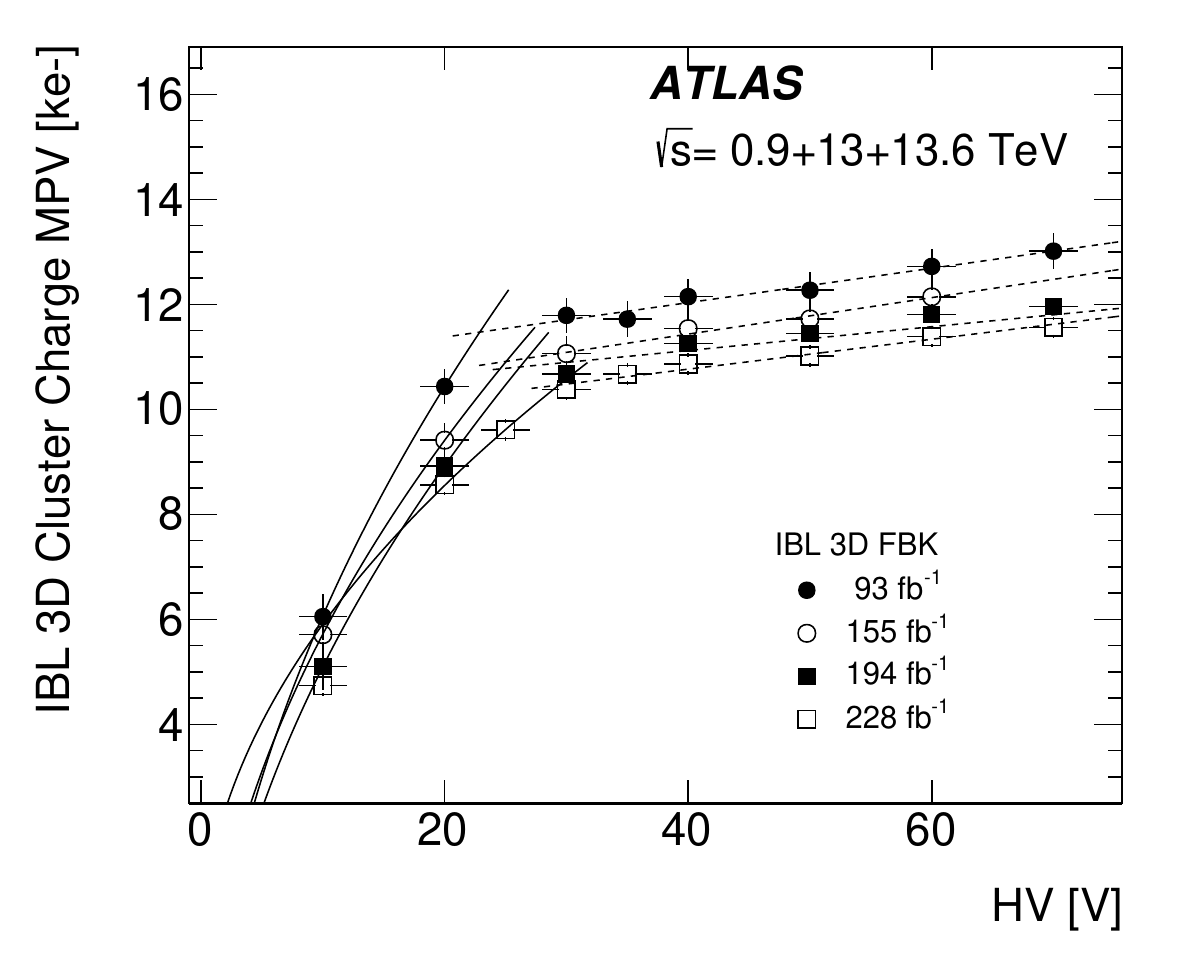}
\label{fig:3DFBKChargeVsHV}
}
\caption{Most probable value (MPV) of the pixel cluster's charge, corrected for particle path length in the Si, for IBL 3D sensors of \protect\subref{fig:3DCNMChargeVsHV} CNM and \protect\subref{fig:3DFBKChargeVsHV} FBK design for clusters associated with reconstructed particle tracks as a function of the applied detector bias voltage (HV). Points represent data recorded during HV scans from early 2017 to the end of 2023, after delivered  integrated luminosities ranging from 93 to 228~fb$^{-1}$, corresponding to values of the average particle fluence on the sensors of $(3.3{-}8)\times 10^{14}$ n-eq~cm$^{-2}$. The \enquote{depletion voltage} is defined by the HV value where the MPV values transition from square root (continuous lines) to linear (dashed lines) behaviour.~\label{fig:3DDesignChargeVsHV}}
\end{figure}
\begin{figure}[hb!]
\centering
\includegraphics[width=0.60\textwidth]{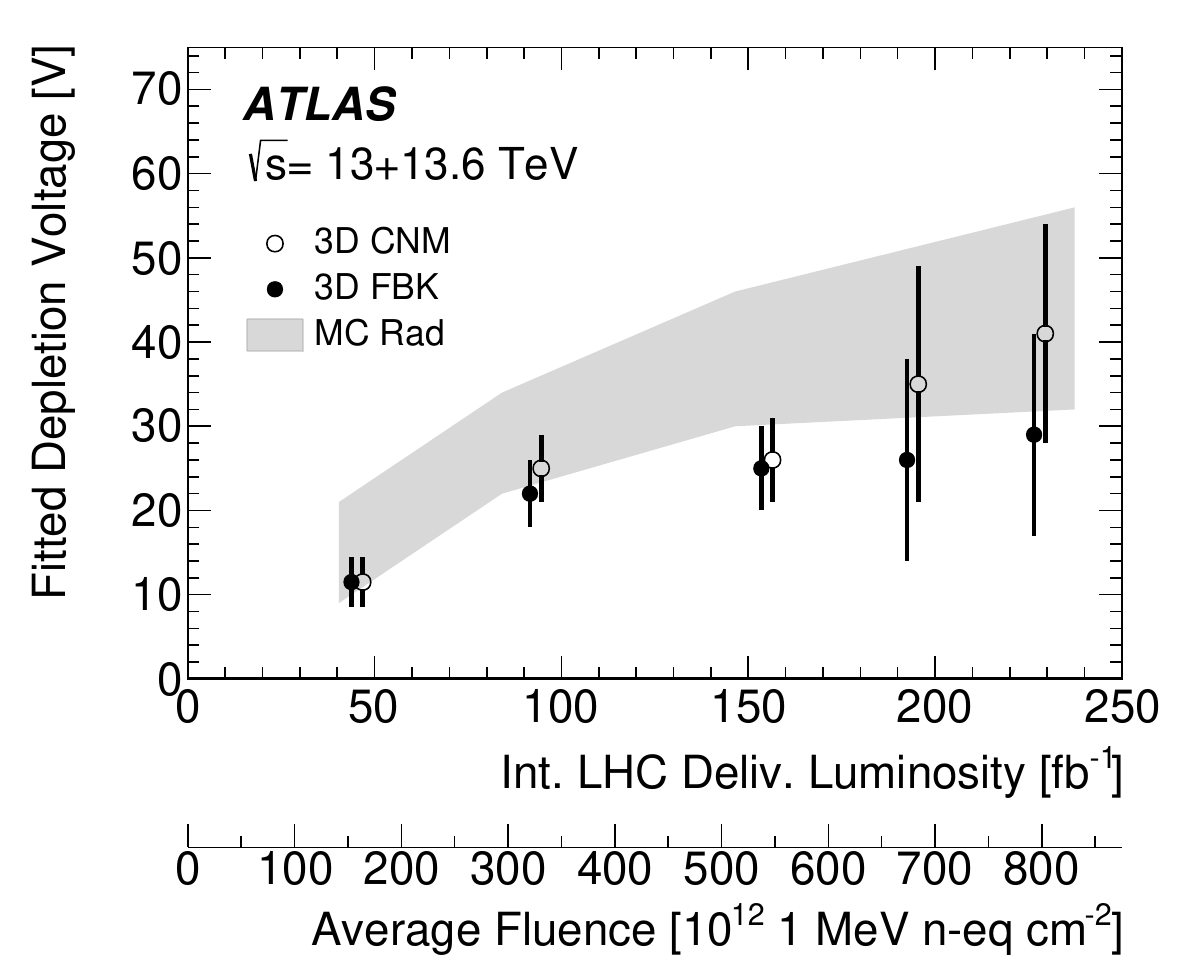}
\caption{Fitted \enquote{depletion voltage} values for 3D FBK and CNM sensors as a function of delivered integrated luminosity and particle fluence in data, compared with simulation predictions. Data points for FBK and CNM are slightly displaced along the fluence axis for readability.~\label{fig:3DDeplVsLum}}
\end{figure}

The charge response vs.\ bias voltage was studied separately for 3D sensors of FBK and CNM design. Results are shown in Figure~\ref{fig:3DDesignChargeVsHV}. The different slopes observed for the MPV charge increase vs.\ HV for FBK and CNM sensors can be ascribed to the different column depths, and was already observed before and after irradiation in the R\&D phase~\cite{DAVIA2014151,DellaBetta:2013owd}. Due to the \enquote{fully-through} columns illustrated in Figure~\ref{fig:3Dschematics}(left), the depleted volume in FBK sensors spreads more rapidly with the applied voltage, so that the increase in the collected charge is initially steeper, whereas the slope of the curve becomes consistently smaller in the regime above depletion. In contrast, in CNM sensors, the silicon regions between the column tips and the opposite surfaces require a higher voltage to become efficient, making the initial increase of the collected charge less steep, and also increasing the slope above the depletion point. The analysis of the HV scans shows a moderate increase of the depletion voltage from 2016 to the end of 2023. The fitted values of the depletion voltage are shown as a function of the integrated luminosity and particle fluence in Figure~\ref{fig:3DDeplVsLum} and compared with simulation predictions.

\subsubsection{Charge collection vs.\ depth of charge generation}
\label{sec:perf-chargedepth}

An important difference between 3D and planar pixel sensors arises from the effect of radiation damage on the charge collection efficiency as a function of the depth of charge generation in the Si bulk. This efficiency is expected to remain constant with depth in 3D sensors after irradiation, due to the specific layout of the pixel electrodes. In planar sensors, however, charge trapping due to Si bulk radiation damage significantly reduces the charge collection efficiency as the depth of charge generation increases. This effect, which is the main source of CCE loss in planar sensors, was measured for planar and 3D sensors using particle tracks from collision data and inclined muon tracks from the cosmic-ray data.

\begin{figure}[hb!]
\centering
\subfloat[]{
\includegraphics[width=0.5\textwidth]{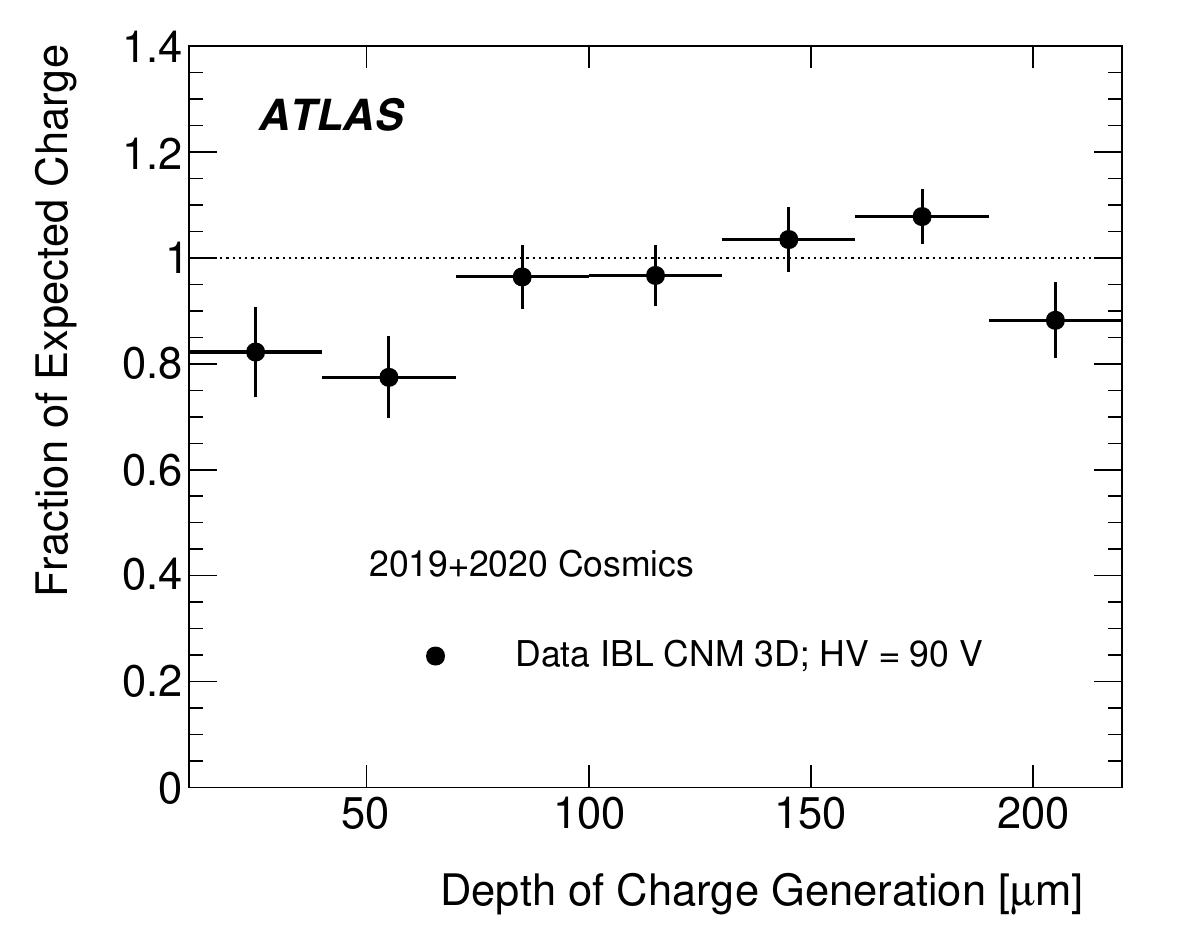}
\label{fig:3DCNMChargeDepth}
}
\subfloat[]{
\includegraphics[width=0.5\textwidth]{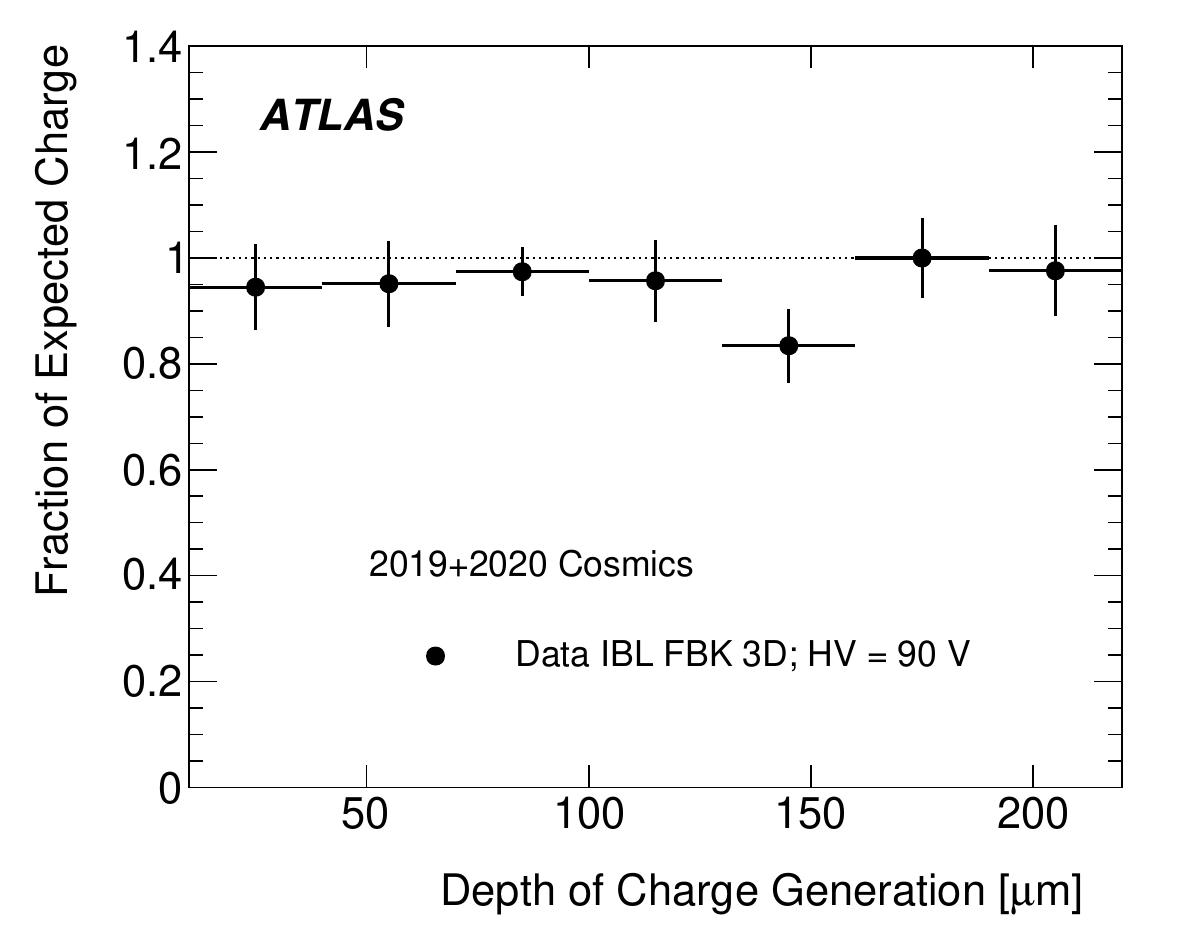}
\label{fig:3DFBKChargeDepth}
}
\caption{Charge collection efficiency as a function of the estimated depth of charge generation after an accumulated dose equivalent to $\sim$160~fb$^-1$ in cosmic-ray muon-track data in IBL 3D \protect\subref{fig:3DCNMChargeDepth} CNM and \protect\subref{fig:3DFBKChargeDepth} FBK sensors. The efficiency is computed by comparing the fraction of the cluster charge deposited on the pixels ordered from the extrapolated point of track entrance in the Si substrate to that of exit in the transverse projection.\label{fig:3DChargeDepth}}
\end{figure}

The measurement method is based on the correlation between the depth of charge generation and the position of each individual pixel in the cluster for tracks traversing the sensitive Si thickness at shallow incidence angles. By comparing the fraction of the total cluster charge collected on each pixel with the fraction of the total track length in the Si under each pixel, it is possible to determine the charge collection efficiency as a function of depth. For this method to be reliable, two conditions must be fulfilled. First, the distance between the points of track entrance and exit must be large compared to the pixel pitch, so that the charge is deposited uniformly below several adjacent pixels. Second, the extrapolated track's position on the detector surface must have a resolution that is small compared to the pixel pitch.

In collision data, these conditions cannot be fulfilled easily for the 3D sensors, since at small incidence angle (in the polar angle $\theta$) the extrapolation resolution deteriorates due to multiple scattering and the detector spacing. Instead, cosmic rays provide a sample of high-momentum muons at shallow incidence in the azimuthal angle $\phi$, where the pixel pitch is smaller. This configuration is well suited for the study of charge collection vs.\ depth. In addition to the selection criteria mentioned above, this study requires that 3D pixel clusters have more than two pixels along the fine pitch ($r{-}\phi$) projection and a single pixel along the long pitch ($z$) projection.

Results are presented in Figure~\ref{fig:3DChargeDepth}, where the average ratio of the measured fraction of the total cluster charge registered on a given pixel to the expected fraction is shown as a function of the average depth of passage of the particle tracks below that pixel. As expected, these distributions are flat within statistical uncertainties, confirming that the charge collection efficiency in 3D sensors is independent of the depth at which the charge is created even after Si bulk radiation damage. For comparison, the results of the same measurement performed with IBL planar sensors using the long pitch projection and 13~TeV collision data, shown in Figure~\ref{fig:Planar_Depth_Charge} for data and radiation damage simulation, show a significant decrease of the charge collection efficiency with the depth of charge generation for irradiated sensors.

\subsection{Spatial resolution}
\label{sec:perf-spatialres}

The spatial resolution of reconstructed hits on 3D pixel sensors was measured in the transverse plane, where the read-out pitch is smaller, using the active region of adjacent module overlaps in $\phi$, using a technique already exploited by ATLAS~\cite{ATL-INDET-PUB-2016-001, ATL-PHYS-PUB-2022-033}.
In these regions, particles traversing the IBL layers generate one hit on each of the two overlapping modules. The spatial resolution can be extracted from the distribution of the corrected difference $\Delta_{r-\phi}$ of the positions of these hits. Corrections are applied to account for geometrical effects due to the different radial positions of the hits and the particle incidence angle. The width of the measured $\Delta$ distribution is $\sqrt{2}$ times larger than the single-hit resolution, assuming that the two hits have the same spatial resolution. This method has the advantage of depending only weakly on the reconstructed track's parameter values and their uncertainties and on detector alignment. In order to reduce the effect of multiple scattering, the analysis is restricted to pixel hits associated with particle tracks with transverse momentum in excess of 4~GeV.

\begin{figure}[ht!]
\centering
\includegraphics[width=0.55\textwidth]{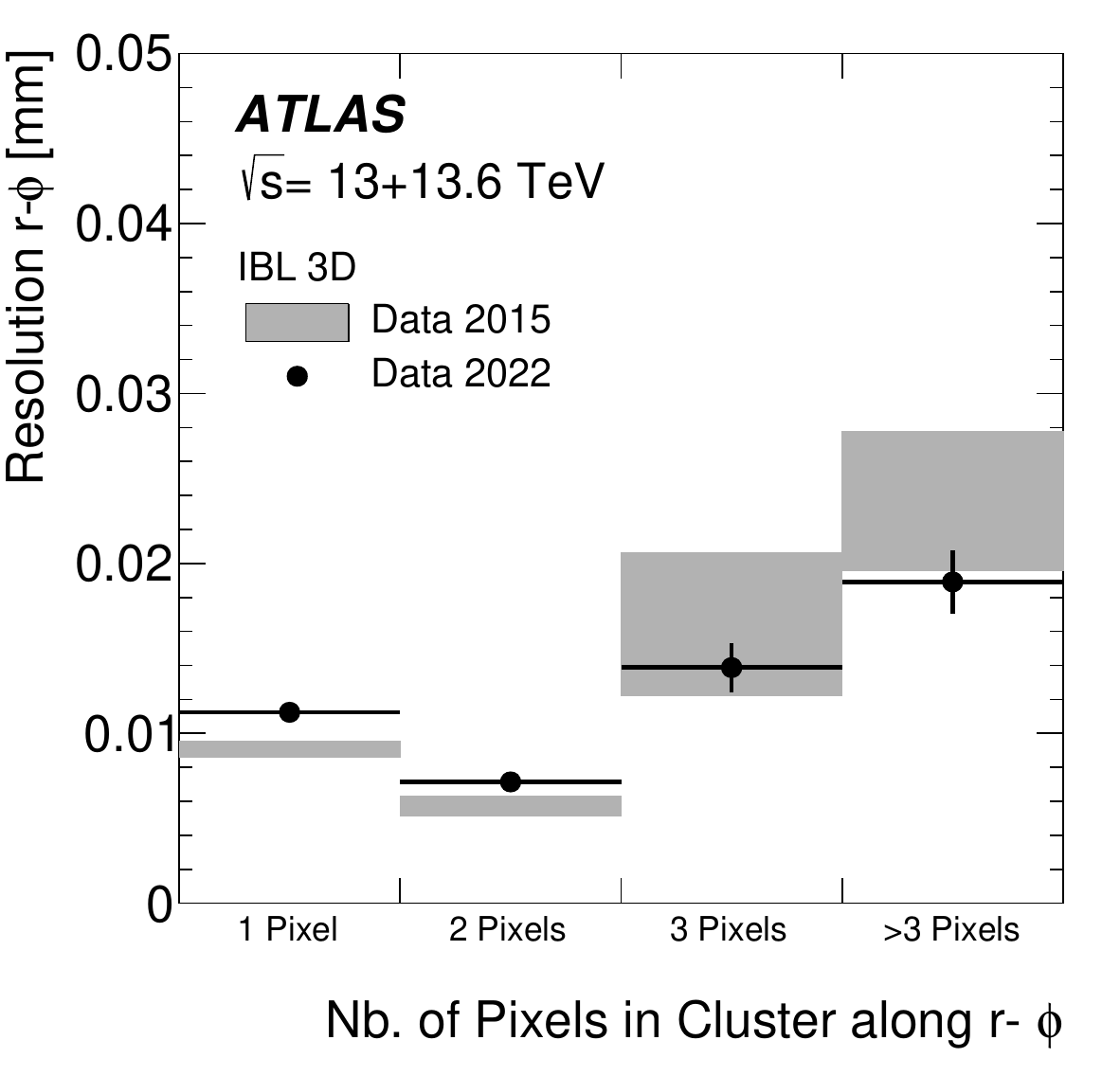}
\caption{Measured IBL 3D hit resolution in the transverse projection as a function of the cluster width along $r{-}\phi$ in $pp$ collision events collected in 2015 (at an average integrated luminosity of 2.5~fb$^{-1}$ ) shown as grey shaded areas and in 2022 (at an integrated luminosity of 172~fb$^{-1}$) shown as filled circles with errors bars. \label{fig:3DResXVsNPix}}
\end{figure}

Averaged over all cluster sizes, the transverse spatial resolutions measured in 2015 and 2022 data are $(8.6\pm0.4)$ and $(9.7\pm0.4)$~$\mu$m, respectively. The predictions from radiation damage simulation for the corresponding fluences are $(8.1\pm0.3)$ and $(9.1\pm0.4)$~$\mu$m, respectively. Their dependences on the number of pixels in the cluster along the $r{-}\phi$ projection are compared in Figure~\ref{fig:3DResXVsNPix}. The poorer measured spatial resolution for clusters with more than two pixels in the transverse projection is due to delta-ray production and cross-talk effects. The evolution of the pixel multiplicity with fluence is shown is Figure~\ref{fig:3DNPix}. Given the large pixel multiplicity in the longitudinal projections, due to the particles' small angle of incidence, the total charge in the 3D pixel clusters is large and radiation damage effects on the spatial resolution correspondingly small.


\section{Conclusions}
\label{sec:conclusions}

%
Pixel sensors in 3D technology have taken LHC collision data at the outer ends of the staves of the innermost layer of the ATLAS Pixel Detector since 2015. Through these years of LHC operation, the 3D sensors have received an average particle fluence of ${\simeq}\,9\times10^{14}$~n-eq~cm$^{-2}$, which is expected to grow to ${\simeq}\,1.5\times10^{15}$~n-eq~cm$^{-2}$ by the end of \RunThr in 2025. Their performance has been studied as a function of integrated luminosity and particle fluence in comparison with radiation damage simulation predictions in order to highlight their tolerance to radiation compared to pixel sensors in planar technology.

The radiation damage simulation based on detailed maps of the electric field in the sensor Si bulk gives a good description of charge collection properties after irradiation. IBL 3D pixels show better response in terms of charge collection efficiency, with a measured $\sim$15\% reduction in the charge collection efficiency loss compared to that of planar sensors at the end of 2023, after a fluence of $8.3\times10^{14}$~n-eq~cm$^{-2}$. This is predicted to grow to $\sim$25\% at the end of \RunThr, with a fluence of $1.5\times10^{15}$~n-eq~cm$^{-2}$.

The analysis of bias voltage scans performed with collision data shows only a moderate increase of the depletion voltage from the beginning of 2016 to the end of 2023. The charge response vs. bias voltage for 3D sensors with \enquote{fully-through} columns shows that the depleted volume spreads more rapidly with the applied voltage and the increase in the collected charge is initially steeper than for sensors with less column depth that require higher voltage to become fully efficient. The charge collection efficiency is found to not depend on the depth of charge generation in the Si bulk even after irradiation, as expected due to the 3D sensor structure. The spatial resolution is measured to be better than 10~$\mu$m with a 50~$\mu$m read-out pitch, after receiving a fluence of ${\simeq}\,5.4\times10^{14}$~n-eq~cm$^{-2}$.

The results reported here confirm the superior radiation hardness of 3D sensors with respect to planar sensors at the LHC. Additionally, the lower high voltage requirements for full sensor depletion require significantly less power consumption. These results are important for projecting the performance of detectors in the HL-LHC era, and for informing research and development for detectors for future experiments.


%
\section*{Acknowledgements}
%

%
%

%
%

We thank CERN for the very successful operation of the LHC and its injectors, as well as the support staff at
CERN and at our institutions worldwide without whom ATLAS could not be operated efficiently.

The crucial computing support from all WLCG partners is acknowledged gratefully, in particular from CERN, the ATLAS Tier-1 facilities at TRIUMF/SFU (Canada), NDGF (Denmark, Norway, Sweden), CC-IN2P3 (France), KIT/GridKA (Germany), INFN-CNAF (Italy), NL-T1 (Netherlands), PIC (Spain), RAL (UK) and BNL (USA), the Tier-2 facilities worldwide and large non-WLCG resource providers. Major contributors of computing resources are listed in Ref.~\cite{ATL-SOFT-PUB-2023-001}.

We gratefully acknowledge the support of ANPCyT, Argentina; YerPhI, Armenia; ARC, Australia; BMWFW and FWF, Austria; ANAS, Azerbaijan; CNPq and FAPESP, Brazil; NSERC, NRC and CFI, Canada; CERN; ANID, Chile; CAS, MOST and NSFC, China; Minciencias, Colombia; MEYS CR, Czech Republic; DNRF and DNSRC, Denmark; IN2P3-CNRS and CEA-DRF/IRFU, France; SRNSFG, Georgia; BMBF, HGF and MPG, Germany; GSRI, Greece; RGC and Hong Kong SAR, China; ISF and Benoziyo Center, Israel; INFN, Italy; MEXT and JSPS, Japan; CNRST, Morocco; NWO, Netherlands; RCN, Norway; MNiSW, Poland; FCT, Portugal; MNE/IFA, Romania; MESTD, Serbia; MSSR, Slovakia; ARIS and MVZI, Slovenia; DSI/NRF, South Africa; MICIU/AEI, Spain; SRC and Wallenberg Foundation, Sweden; SERI, SNSF and Cantons of Bern and Geneva, Switzerland; NSTC, Taipei; TENMAK, T\"urkiye; STFC/UKRI, United Kingdom; DOE and NSF, United States of America.

Individual groups and members have received support from BCKDF, CANARIE, CRC and DRAC, Canada; CERN-CZ, FORTE and PRIMUS, Czech Republic; COST, ERC, ERDF, Horizon 2020, ICSC-NextGenerationEU and Marie Sk{\l}odowska-Curie Actions, European Union; Investissements d'Avenir Labex, Investissements d'Avenir Idex and ANR, France; DFG and AvH Foundation, Germany; Herakleitos, Thales and Aristeia programmes co-financed by EU-ESF and the Greek NSRF, Greece; BSF-NSF and MINERVA, Israel; NCN and NAWA, Poland; La Caixa Banking Foundation, CERCA Programme Generalitat de Catalunya and PROMETEO and GenT Programmes Generalitat Valenciana, Spain; G\"{o}ran Gustafssons Stiftelse, Sweden; The Royal Society and Leverhulme Trust, United Kingdom.

In addition, individual members wish to acknowledge support from Armenia: Yerevan Physics Institute (FAPERJ); CERN: European Organization for Nuclear Research (CERN PJAS); Chile: Agencia Nacional de Investigaci\'on y Desarrollo (FONDECYT 1230812, FONDECYT 1230987, FONDECYT 1240864); China: Chinese Ministry of Science and Technology (MOST-2023YFA1605700), National Natural Science Foundation of China (NSFC - 12175119, NSFC 12275265, NSFC-12075060); Czech Republic: Czech Science Foundation (GACR - 24-11373S), Ministry of Education Youth and Sports (FORTE CZ.02.01.01/00/22\_008/0004632), PRIMUS Research Programme (PRIMUS/21/SCI/017); EU: H2020 European Research Council (ERC - 101002463); European Union: European Research Council (ERC - 948254, ERC 101089007), Horizon 2020 Framework Programme (MUCCA - CHIST-ERA-19-XAI-00), European Union, Future Artificial Intelligence Research (FAIR-NextGenerationEU PE00000013), Italian Center for High Performance Computing, Big Data and Quantum Computing (ICSC, NextGenerationEU); France: Agence Nationale de la Recherche (ANR-20-CE31-0013, ANR-21-CE31-0013, ANR-21-CE31-0022, ANR-22-EDIR-0002), Investissements d'Avenir Labex (ANR-11-LABX-0012); Germany: Baden-Württemberg Stiftung (BW Stiftung-Postdoc Eliteprogramme), Deutsche Forschungsgemeinschaft (DFG - 469666862, DFG - CR 312/5-2); Italy: Istituto Nazionale di Fisica Nucleare (ICSC, NextGenerationEU), Ministero dell'Università e della Ricerca (PRIN - 20223N7F8K - PNRR M4.C2.1.1); Japan: Japan Society for the Promotion of Science (JSPS KAKENHI JP22H01227, JSPS KAKENHI JP22H04944, JSPS KAKENHI JP22KK0227, JSPS KAKENHI JP23KK0245); Netherlands: Netherlands Organisation for Scientific Research (NWO Veni 2020 - VI.Veni.202.179); Norway: Research Council of Norway (RCN-314472); Poland: Ministry of Science and Higher Education (IDUB AGH, POB8, D4 no 9722), Polish National Agency for Academic Exchange (PPN/PPO/2020/1/00002/U/00001), Polish National Science Centre (NCN 2021/42/E/ST2/00350, NCN OPUS nr 2022/47/B/ST2/03059, NCN UMO-2019/34/E/ST2/00393, UMO-2020/37/B/ST2/01043, UMO-2021/40/C/ST2/00187, UMO-2022/47/O/ST2/00148, UMO-2023/49/B/ST2/04085); Slovenia: Slovenian Research Agency (ARIS grant J1-3010); Spain: Generalitat Valenciana (Artemisa, FEDER, IDIFEDER/2018/048), Ministry of Science and Innovation (MCIN \& NextGenEU PCI2022-135018-2, MICIN \& FEDER PID2021-125273NB, RYC2019-028510-I, RYC2020-030254-I, RYC2021-031273-I, RYC2022-038164-I), PROMETEO and GenT Programmes Generalitat Valenciana (CIDEGENT/2019/027); Sweden: Swedish Research Council (Swedish Research Council 2023-04654, VR 2018-00482, VR 2022-03845, VR 2022-04683, VR 2023-03403, VR grant 2021-03651), Knut and Alice Wallenberg Foundation (KAW 2018.0157, KAW 2018.0458, KAW 2019.0447, KAW 2022.0358); Switzerland: Swiss National Science Foundation (SNSF - PCEFP2\_194658); United Kingdom: Leverhulme Trust (Leverhulme Trust RPG-2020-004), Royal Society (NIF-R1-231091); United States of America: U.S. Department of Energy (ECA DE-AC02-76SF00515), Neubauer Family Foundation.

%
%


%
\clearpage
\printbibliography
\clearpage
\input{atlas_authlist}
%

%
%
%
%

%
%
%
%

\end{document}

%% file: atlas_authlist.tex
 
\begin{flushleft}
\hypersetup{urlcolor=black}
{\Large The ATLAS Collaboration}

\bigskip

\AtlasOrcid[0000-0002-6665-4934]{G.~Aad}$^\textrm{\scriptsize 104}$,
\AtlasOrcid[0000-0001-7616-1554]{E.~Aakvaag}$^\textrm{\scriptsize 17}$,
\AtlasOrcid[0000-0002-5888-2734]{B.~Abbott}$^\textrm{\scriptsize 123}$,
\AtlasOrcid[0000-0002-0287-5869]{S.~Abdelhameed}$^\textrm{\scriptsize 119a}$,
\AtlasOrcid[0000-0002-1002-1652]{K.~Abeling}$^\textrm{\scriptsize 56}$,
\AtlasOrcid[0000-0001-5763-2760]{N.J.~Abicht}$^\textrm{\scriptsize 50}$,
\AtlasOrcid[0000-0002-8496-9294]{S.H.~Abidi}$^\textrm{\scriptsize 30}$,
\AtlasOrcid[0009-0003-6578-220X]{M.~Aboelela}$^\textrm{\scriptsize 45}$,
\AtlasOrcid[0000-0002-9987-2292]{A.~Aboulhorma}$^\textrm{\scriptsize 36e}$,
\AtlasOrcid[0000-0001-5329-6640]{H.~Abramowicz}$^\textrm{\scriptsize 154}$
% ...
\bigskip
\\

% ...
$^{17}$Department for Physics and Technology, University of Bergen, Bergen; Norway.\\
% ...
$^{30}$Physics Department, Brookhaven National Laboratory, Upton NY; United States of America.\\
% ...
$^{36}$$^{(a)}$Facult\'e des Sciences Ain Chock, Universit\'e Hassan II de Casablanca;$^{(b)}$Facult\'{e} des Sciences, Universit\'{e} Ibn-Tofail, K\'{e}nitra;$^{(c)}$Facult\'e des Sciences Semlalia, Universit\'e Cadi Ayyad, LPHEA-Marrakech;$^{(d)}$LPMR, Facult\'e des Sciences, Universit\'e Mohamed Premier, Oujda;$^{(e)}$Facult\'e des sciences, Universit\'e Mohammed V, Rabat;$^{(f)}$Institute of Applied Physics, Mohammed VI Polytechnic University, Ben Guerir; Morocco.\\
% ...
$^{45}$Physics Department, Southern Methodist University, Dallas TX; United States of America.\\
% ...
$^{50}$Fakult\"{a}t Physik , Technische Universit{\"a}t Dortmund, Dortmund; Germany.\\
% ...
$^{56}$II. Physikalisches Institut, Georg-August-Universit\"{a}t G\"ottingen, G\"ottingen; Germany.\\
% ...
$^{104}$CPPM, Aix-Marseille Universit\'e, CNRS/IN2P3, Marseille; France.\\
% ...
$^{119}$$^{(a)}$New York University Abu Dhabi, Abu Dhabi;$^{(b)}$United Arab Emirates University, Al Ain; United Arab Emirates.\\
% ...
$^{123}$Homer L. Dodge Department of Physics and Astronomy, University of Oklahoma, Norman OK; United States of America.\\
% ...
$^{154}$Raymond and Beverly Sackler School of Physics and Astronomy, Tel Aviv University, Tel Aviv; Israel.\\
% ...
$^{a}$ Also Affiliated with an institute covered by a cooperation agreement with CERN.\\
$^{b}$ Also at An-Najah National University, Nablus; Palestine.\\
$^{c}$ Also at Borough of Manhattan Community College, City University of New York, New York NY; United States of America.\\
$^{d}$ Also at Center for Interdisciplinary Research and Innovation (CIRI-AUTH), Thessaloniki; Greece.\\
$^{e}$ Also at Centro Studi e Ricerche Enrico Fermi; Italy.\\
$^{f}$ Also at CERN, Geneva; Switzerland.\\
% ...

\end{flushleft}
